\documentclass[a4paper]{article}

\usepackage{graphics}
\usepackage{amsmath}
\usepackage{amssymb}
\usepackage{amscd}
\usepackage[ruled]{algorithm2e}

\newcommand{\tr}[1]{\ensuremath{\stackrel{#1}{\longrightarrow}}}
\newcommand{\I}[1]{\ensuremath{I_{#1}}}
\newcommand{\C}[1]{\ensuremath{C_{#1}}}
\newcommand{\spmn}{\ensuremath{SPM(n)}}
\newcommand{\spmnp}{\ensuremath{SPM(n+1)}}
\newcommand{\spm}{\textsc{spm}}
\newcommand{\soc}{\textsc{soc}}
\newcommand{\cfg}{\textsc{cfg}}
\newcommand{\proof}{\textbf{Proof: }}
\newcommand{\cqfd}{\hfill \fbox{} \vskip 0.2cm}
\newcommand{\fb}[1]{\ensuremath{^{\downarrow_{#1}}}}
\newcommand{\inc}[2]{\ensuremath{#1\fb{#2}}}

\newcommand{\di}[2]{\ensuremath{_{d_{#1}=#2}}}

\newcommand{\spmi}{\ensuremath{SPM(\infty)}}
\newcommand{\spt}{\ensuremath{SPT(\infty)}}

\newcommand{\fig}[2]{\begin{figure}[!h] \centerline{\includegraphics{#1.eps}} \caption{\label{fig_#1} #2} \end{figure}}

\title{\textbf{Structure of some sand piles model}\\
\large{Matthieu Latapy, Roberto Mantaci,\\
Michel Morvan and Ha Duong Phan}\\
November 1998}
\date{}

\begin{document}
\maketitle

\newtheorem{prop}{Proposition}
\newtheorem{theo}{Theorem}
\newtheorem{corol}{Corollary}
\newtheorem{lemme}{Lemma}
\newtheorem{definition}{Definition}
\newtheorem{notation}{Notation}
\newtheorem{rem}{Remark}

\vskip -1cm
\textbf{Abstract:} \spm\ (Sand Pile Model) is a simple discrete dynamical
system used in physics to represent granular objects. It is deeply related to
integer partitions, and many other combinatorics problems, such as tilings
or rewriting systems. The evolution of the system started
with $n$ stacked grains generates a lattice, denoted by \spmn.
We study here the structure of this lattice. We first explain how it can be
constructed, by showing its strong self-similarity property. Then, we define \spmi, a
natural extension of \spm\ when one starts with an infinite number of grains.
Again, we give an efficient
construction algorithm and a coding of this lattice using a self-similar tree.
The two approaches give different recursive 
formulae for $|\spmn|$.


\vskip 0.5cm

\section {Introduction}

\subsection{Motivations and context}

In 1987, Bak, Tang and Wiesenfeld \cite{BTWInitial} introduced the
important notion of \textbf{self-organisation criticality
(\soc)}: when certain systems in a steady
state (named critical state) are slightly perturbated, they evolve back to
another steady state.
This evolution implies some arbitrarily high modifications of the system.

The typical example is an avalanche on a sand pile. At first, the pile 
is in a steady state and the perturbation consists in adding a grain on the pile.
As a consequence, the pile evolves to a new steady state, with an avalanche starting where the
grain was dropped.
The fact that this avalanche size may be arbitrarily high
is the main characteristic of
\soc\ systems.

Since the appearance of this paper, many physicists and biologists have recognized
these properties in natural systems, and the \soc\ family
still grows
(see \cite{LivreSOC}, \cite{SOC} for example); many publications
on this topic appeared recently
\cite{LivreSOC}, \cite{LivreFr}, \cite{LivreBak},
\cite{LivreFract}.
These phenomena are of particular interest in surface grow studies
\cite{LivreLaurent}, in geophysics
\cite{TremblementsDeTerre}, in plasma confinment, in astrophysics,
and many other, including,
of course, studies of
granular systems like dunes \cite{LivreFr} and molecule agregation \cite{LivreLaurent}.

The essence of these phenomena is captured by a well known model in
game theory and combinatorics, the
\textbf{Chip Firing Game}
(\cfg).
The most general notion of \cfg\ is a directed graph
$G=(V,E)$ where a threshold $\delta _{v}$
and a load $l(v)$ are given to each vertex $v$.
Intuitively, $l(v)$ represents the number of chips stored
at $v$. The game evolves with respect to the following rule:
if $v\in V$ contains more than
$\delta _v$ chips, then it gives $\delta _v$ of them to its neighbours,
i.e. the load of the vertex $v$ is decreased by $\delta _v$ and the load of each of
its neighbours is increased by
$\frac{\delta _v}{n_v}$ where $n_v$ is the number of $v$ neighbours.
In general, one takes $\delta _v = n_v$, but
$\delta _v = \infty$ if $n_v = 0$ ($v$ is then called a sink).
See Figure~\ref{fig_cfg} for an example of such a \cfg.

\fig{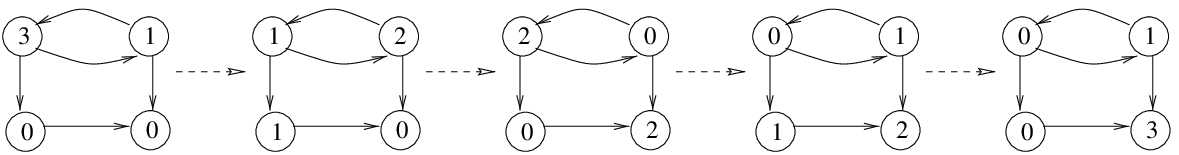}{An example of \cfg.
Each vertex load is written in this vertex.}

Under certains conditions, the \cfg\ converges to a steady state
(see for example \cite{TheseEriksson}).
The addition of one chip on a vertex
$v \in V$ when the system is in a steady state causes a redistribution
of the chips. During this redistribution, an arbitrary number of vertexs
may be concerned. To give an example, we can consider the case where
$l(\nu)=0$ and $l(v)=\delta
_v\ \forall\ v\ \in\ V \setminus \lbrace \nu \rbrace$.
If one adds successively
$\delta_{\nu} -1$ chips on $\nu$, the only concerned vertex is $\nu$ (the
system remains steady). If one adds one more grain, every vertex will
be concerned (if $G$ is connected). Such a
diffusion can be arbitrary large \cite{Univ_cfg},
depending on the initial state of the
system, and is always started by addition of one grain.
Such a propagation is called an \emph{avalanche}.

A particular case of this model is widely studied: the sand pile
model on a rectangular grid\,\footnote{The standard term is \emph{lattice} but,
since we will use orders theory in the following, where the word \emph{lattice}
takes another meaning, we use here the word \emph{grid}.}.
The graph $G$ in this case is undirected. It is a rectangular finite lattice
and the value of $\delta _v$ and of $n_v$ is $4$ for all vertex $v$ except one singular vertex $v$ which is
linked once to any vertex on the border of the lattice and twice to the four
corners, and such that
$\delta _v = \infty$.
The distinguished vertex acts like a sink: it never gives away any
of its grains and could be considered as collecting the grains that
leaves the system. If the load of a vertex inside the lattice is more
than $4$, then it gives one grain to each of its four neighbours
(see Figure~\ref{fig_tas} for an example where the distinguished vertex is not
represented since its load does not influence the evolution of the system).
This is the model deeply studied by Dhar
\cite{Dhar1}, \cite{Dhar2}. In particular, one can show that adding a grain
turns the system into an unsteady state, and that after auto-reorganisation
it reaches a new steady state \cite{These_Marguin}.
This confirms that we are in the
\soc\ context.
Cori and Rossin \cite{Cori} generalized this notion to any rooted graph and obtained similar
results.

\fig{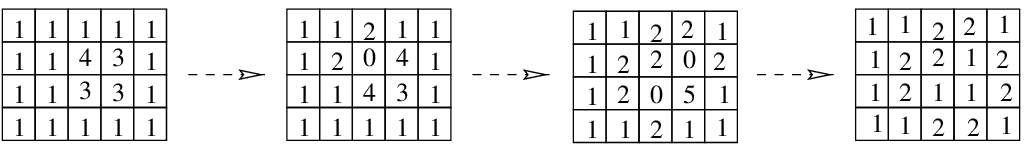}{Example of (parallel) sand pile.}

Another special case of \cfg\ 
is the \textbf{Sand Piles Model} (\spm). The graph $G$ in this case is an undirected
chain, infinite on the right:
$V = \mathbb{N}$, $E=\lbrace (i,i+1)\ \forall\ i \in
\mathbb{N} \rbrace$, $\delta _v = 2\mbox{ for all }v > 0$ and $\delta_0 = \infty$
(see Figure~\ref{fig_equiv}(a)).
This model is equivalent to the following. Consider an infinite chain
of columns, each containing a vertical pile of grains.
The height difference between the column $c_i$ and its right neighbour column $c_{i+1}$
is denoted by
$d(i)$.
If $d(i)$ is greater than or equal to $2$ then a grain falls down from $c_i$ to $c_{i+1}$
(see Figure~\ref{fig_equiv}(b)).
If $i>0$, we call $c_{i-1}$ the left neighbour of $c_i$ and then $d(c_{i-1})$ and $d(c_{i+1})$
are increased by $1$ while $d(c_i)$ is decreased by $2$. We find again
our initial definition of \spm, with a coding of the pile by
height differences.

Notice that \spm\ is less general than the lattice sand pile
but is not a particular case of it: the number of vertices  $v$ with
$l(v)\not= 0$ is not bounded
in \spm, and it effectively grows with the number of grains.
Moreover, the model \spm\ has no sink, which is a fundamental difference.
If we represent a pile by the
$t$-uple of its columns height, each configuration of the pile
represents a partition of the total number of grains
(see Figure~\ref{fig_equiv}(c)).

\fig{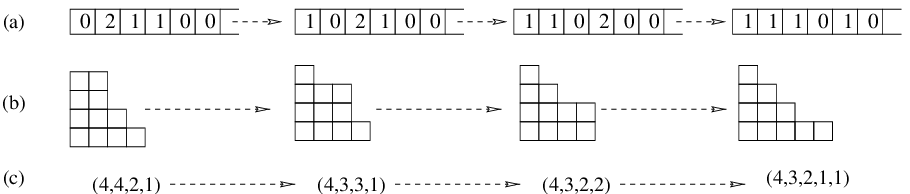}{Three ways to see \spm.}

In computer science, the \cfg\ models several problems and is applied in several algorithms (see for example \cite{Goles_GrArbre}).
\spm\ itself admits natural interpretations in algorithmic terms. We
give here two examples about dynamical distribution of jobs on a processors
network \cite{Reparti} \cite{GameOfCards} \cite{HDP1}.
Each column of a sand pile represents a processor,
a grain represents a job. One can imagine the processors are
connected on a ring (like Token Ring):
each processor can only communicate directly with
its right neighbour. It corresponds to the move of a grain from one column to
another.
Since only neighbour processors can communicate, the communications can be
processed in parallel and the parallel \spm\ is a good model for this problem
\cite{TheseJD}. If on the contrary the communication medium
is a shared bus (like Ethernet or certain multiprocessors), we can study
the evolution of sequential
\spm\ to avoid collisions.

In the following, we are going to discuss some lattice properties of the
above dynamical systems. Let us recall that a lattice can be described
as a partial order such that two elements $a$ and $b$ admit a least
upper bound (called supremum of $a$ and $b$ and denoted by $\sup(a,b)$)
and a greatest lower bound (called infimum of $a$ and $b$ and denoted by
$\inf(a,b)$). The element $\sup(a,b)$ is the smallest element among the
elements greater than both $a$ and $b$. The element $\inf(a,b)$ is
defined similarly. A useful result about finite lattices is that a partial
order is a lattice if and only if it admits a greatest element, and any two
elements admit a greatest lower bound.
For more details, see for example \cite{Intro}.
The fact that a dynamical system has the lattice property implies some
important properties, such as convergence.

%
%
%
%
%
%
\subsection{Our model: known results}

Our model is the standard sequential \spm; it consists of an infinite number of
ordered columns, each containing a certain
number of grains. Only the first $k$
columns are non-empty, so the state of the system is described by the $k$-uple
$s=(s_1,s_2,\dots ,s_k)$ where $s_i$ is the number of grains in the column
$i$ for $1\le i\le k$.

The system is initially in the state $N = (n)$. This means that
all the grains are in the first column. At each step, the system evolves with
respect to the following rule: one grain can fall down from column $i$
to column $i+1$ if and only if $s_{i+1}-s_i \ge 2$.
This rule defines a covering relation on the set of reachable
configurations. The reflexive and transitive closure of this relation is an
order, called the dominance order \cite{HDP3}. The set of reachable
configurations from the partition $(n)$ with this order is then a lattice denoted by \spmn\ \cite{HDP3}.

Let $s = (s_1,\dots s_k)$ be a sand pile, the 
\emph{height difference of $s$ at $i$},
denoted by $d_i(s)$, is the integer $s_i - s_{i+1}$
(with the assumption that $s_{k+1} = 0$).
We will say that $s$ has a \emph{step} (resp. \emph{plateau},
resp. \emph{cliff}) at $i$ if and only if
its height difference at $i$ is $1$ (resp. $0$, resp. $\ge 2$).
We extend these definitions by saying that $s$ has stairs (resp. a plateau)
at the interval $[i,j]$ if and only if $s$ has a step  (resp. plateau)
at $k$ for all $i\le k\le j$.
The integer $j-i+1$ is called the \emph{length} of the stairs (resp. plateau).
See Figure~\ref{fig_defs} for examples.

\fig{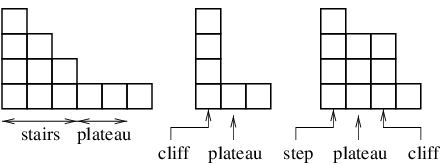}{Examples for the definitions.}

The evolution rule of a sand pile $s=(s_1,\dots ,s_i,s_{i+1},\dots , s_k)$
is then:
one grain can fall from one column to the column on its right if and only if it is at the top of a cliff.
Such a
transition is denoted by $\tr{i}$ where $i$ is the
number of the column from which the grain falls. The sand pile $s'$ is called
a \emph{successor} of $s$,
and $Succ(s)$ denotes the set of all successors of $s$:
$$Succ(s) = \lbrace s' | s \tr{i} s', i\in\mathbb{N} \rbrace.$$
See Figure
\ref{fig_transucc} for an example.

\fig{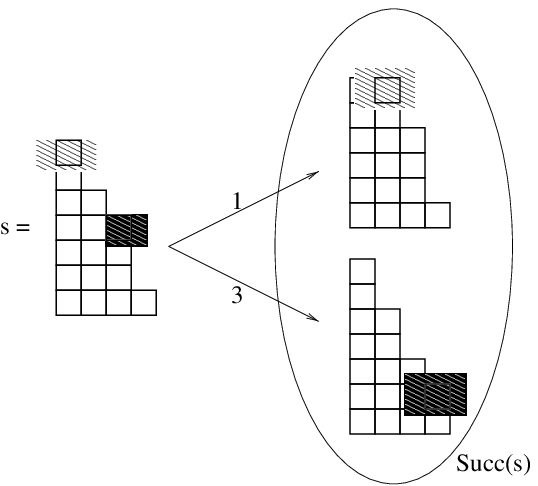}{Transitions, successors.}

Let us now introduce a few notations.
If $s$ is a partition of $n$
then \inc{s}{i}\ is the partition of $n+1$ obtained by adding
a grain on the $i$-th column of $s$ (if it is a partition).
In other words, if $s = (s_1,\dots,s_i,\dots ,s_k)$ then
$\inc{s}{i} = (s_1,\dots ,s_i+1,\dots ,s_k)$.
We also define $\inc{S}{i} = \lbrace \inc{s}{i} | s \in S \rbrace$, where
$S$ is a set of partitions.

We will denote by $e(s)$ the largest integer such that $s$ has stairs at
$[1,e(s)]$. We also define
$P_i$ as the set of the sand piles in \spmn\ that begins
with stairs of length (at least) $i$. In other words, $P_i = \lbrace
s \in \spmn\ |\ e(s)\ge i \rbrace$.

Charaterisations of the fixed point of the system, the minimum element of
the corresponding lattice, and of its
elements are also known:
\begin{theo}[\cite{Goles1}]
\label{CaractPtFixe}
The fixed point of \spmn\ is:
$$S_0 = (k,k-1,\dots,p+1,p,p,p-1,\dots,2,1) $$
where $k$ is the maximal integer such that $S_0$ is a sand pile of $n$
grains, i.e. $k$ is the integer such that
$\frac{k(k+1)}{2} \le n \le \frac{(k+1)(k+2)}{2}$.
\end{theo}
\begin{theo}[\cite{HDP3}]
\label{CaractElts}
A partition $s$ belongs to
\spmn\ if and only if:
\begin{itemize}
\item $s$ does not contain any sequence $p,p,p$ or
      $p,p,(p-1),(p-1)$.
\item there is at least one cliff between two consecutive
      sequences $p,p$ and $q,q$.
\end{itemize}
\end{theo}

In this paper, we study the structure of \spmn. In particular,
we show in the next section how \spmnp\ can be constructed
from \spmn, thus we obtain an algorithm that constructs \spmn\ for
any integer $n$. Afterwards, we define a natural infinite extension,
\spmi, when the system is started with an infinite column of grains.
The study of the structure of \spmi\ permits more
remarks on the self-similarity of the set.
During this study, we obtain interesting recursive formula for
$|\spmn|$.


\section{From \spmn\ to \spmnp}

The goal of this section is the construction of the lattice \spmnp\ from \spmn.
We will construct the graph of the transitive reduction of the lattice, i.e.
the graph of its order relation, without the reflexive edges
($x \longrightarrow x$) and the transitive ones ($x \longrightarrow z$
when $x \longrightarrow y$ and $y \longrightarrow z$).
Each edge of this graph is equivalent to a transition of the \spm\ system.
Therefore, we will label the edge $s \tr{i} s'$ with the number $i$ of
the column of $s$ from which the grain falls in order to obtain $s'$.
We will call the obtained labelled graph the \emph{diagram} of the lattice.
We first give some preliminary results,
then we notice that \spmn\ is a good starting point to construct \spmnp,
and we give a method to obtain \spmnp\ from \spmn.
Finally, we inspect more deeply
the construction algorithm and show a strong self-similarity
in each lattice \spmn. This similarity induces a first recursive formula
for the cardinality of \spmn.

\subsection{Preliminaries}

Let us study what happens when we add one grain on the
$i$-th column of a sand pile $s=(s_1,\dots ,s_i,\dots ,s_k)$
such that $e(s) \ge i-1$.
We obtain the sand pile
$\inc{s}{i} = (s_1,\dots
,s_i+1,\dots ,s_k)$. We want to determine all the possible transitions
from this partition, knowing the possible ones from $s$.
Three cases are possible (as shown in Figure \ref{fig_prelim})
corresponding to the three following propositions.
Recall that we only consider sand piles $s$ with $e(s) \ge i-1$,
since it will be the case of interest for the rest of the paper.

\fig{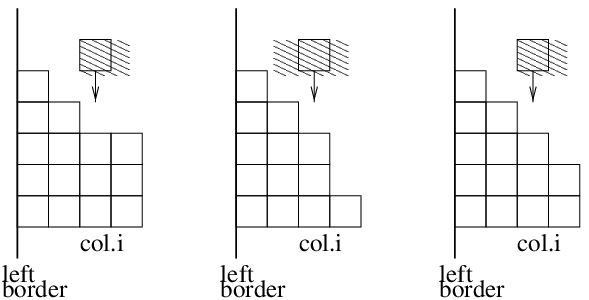}{The three cases considered.}

\begin{prop}[plateau]
 \label{Prop_plateau}
 Let $s \in \spmn$ such that $e(s) \ge i-1$.
 If $s$ has a plateau at $i$ then the possible transitions from
 \inc{s}{i}\ are the same as   the possible transitions from $s$. Moreover, if
 $s \tr{j} t$ then \inc{s}{i} \tr{j} \inc{t}{i}.
 In other words, $Succ(\inc{s}{i}) = \inc{(Succ(s))}{i}$
 and the corresponding edges of
 the diagrams have the same labels.
\end{prop}
\proof
A transition \tr{i} is only possible if there is a cliff at the column $i$.
Now, the set of the columns where $s$ has a cliff is equal to the set of the
columns where \inc{s}{1} has a cliff.
\cqfd

\begin{prop}[cliff]
 Let $s \in \spmn$ such that $e(s) \ge i-1$.
 \label{Prop_falaise}
 If $s$ has a cliff at $i$ then
 \begin{enumerate}
  \item The possible transitions from \inc{s}{i}\ are the same as  
        the possible transitions from $s$
        and if $s \tr{j} t$ then \inc{s}{i} \tr{j} \inc{t}{i}.
        In other words, $Succ(\inc{s}{i}) = \inc{(Succ(s))}{i}$
        and the corresponding edges of
        the diagrams have the same labels.
  \item Moreover, if \inc{s}{i} \tr{i} $s'$ then
        \inc{s}{1} \raisebox{0.2cm}{$\underrightarrow{\mbox{\scriptsize $i\cdot i-1\cdot\ 
	\dots\ \cdot 2\cdot 1 $}}$} $s'$.
        In other words, $s'$ is reachable in \inc{\spmn}{1}\ from
	\inc{s}{1}\ via a path
        labelled $i\cdot i-1\cdot\ \dots\ \cdot 2\cdot 1$.
 \end{enumerate}
\end{prop}
\proof
 \begin{enumerate}
  \item 
A transition \tr{i} is only possible if there is a cliff at the column $i$.
Now, the set of the columns where $s$ has a cliff is equal to the set of the
columns where \inc{s}{1} has a cliff.
  \item The sand pile $s'$ is equal to $\inc{s}{i+1}$ and by
        hypothesis $s$ has a cliff at $i$ and stairs at
        $[1,i-1]$. Therefore, in the sand pile \inc{s}{1}\ a grain can fall
	from the column $i$.
	With this grain's fall, we create a cliff at $i-1$,
	therefore a new grain can now fall from column $i-2$. This process can
	be iterated to obtain $s'$ at the end.
	To sum up, we can write:\\
  $\inc{s}{1} = (s_1+1,\dots ,s_{i-1},s_j,s_{i+1},\dots, s_k)\\
   \tr{i}
   (s_1+1,\dots ,s_{i-1},s_i-1,s_{i+1}+1,\dots, s_k)\\
   \tr{i-1}
   \dots
   \tr{2}
   (s_1+1,s_2-1,\dots ,s_i,s_{i+1}+1,\dots, s_k)\\
   \tr{1}
   (s_1+1-1,s_2,\dots ,s_i,s_{i+1}+1,\dots, s_k) = \inc{s}{i+1} = s'$\\
   It is obvious that all the partitions on this path belong to
   \inc{\spmn}{1}.
 \end{enumerate} \cqfd

\begin{prop}[step]
 Let $s \in \spmn$ such that $e(s) \ge i-1$.
 \label{Prop_escalier}
 If $s$ has a step  at $i$ then the possible transitions from \inc{s}{i}\ 
 are the same as   from $s$ with an additionnal transition on the column $i$:
 \inc{s}{i} \tr{i} \inc{s}{i+1}.
\end{prop}
\proof
The set of columns where \inc{s}{i} has a cliff is equal to the union of
$\lbrace i \rbrace$ and  the set of columns where
$s$ has a cliff.
\cqfd

\subsection{Construction}

Using the preliminary results from the previous section, we will here
obtain an algorithm for the construction
of \spmnp\ from \spmn.
We first show that \spmn\fb{1}\ is a good starting point for the construction
of \spmnp. Recall that \inc{\spmn}{1} is the set of partitions obtained by addition
of one grain on the first column of each partition in \spmn.
Afterwards, we will use the previous propositions to add the
missing elements and transitions in order to complete
\spmn\fb{1}\ into \spmnp.
\label{section_deb}

\begin{prop}
\label{prop_sub}
\spmn\fb{1}\ is a sublattice of \spmnp.
\end{prop}
\proof 
Let us recall that if $a$ and $b$ are two partitions of \spmn\ for
a given $n$, then
$\inf(a,b)$ is their first common descendant and $\sup(a,b)$ is their first
common ancestor. To prove the claim, we must show that:
\begin{itemize}
\item If $\inf(a,b)=c$ is in \spmn\ then $\inf(\inc{a}{1},\inc{b}{1}) = \inc{c}{1}$ is in \spmnp.
\item If $\sup(a,b)=c$ is in \spmn\ then $\sup(\inc{a}{1},\inc{b}{1}) = \inc{c}{1}$ is in \spmnp.
\end{itemize}
Recall that \cite{HDP3}:
$$
\inf(a,b) = c\mbox{ iff for all $j$ one has }\sum_{i=1}^j c_i = min(\sum_{i=1}^j a_i,
\sum_{i=1}^j b_i).
$$
This implies that:
$$
\sum_{i=1}^j c_i+1 = min(\sum_{i=1}^j a_i+1, \sum_{i=1}^j b_i+1) \mbox{ for
all $j$,}
$$
i.e. $c^{\downarrow_1}$ is in \spmnp.

Let now $c$ be equal to $\sup(a,b)$ (in \spmn) and $d$ be equal to
$\sup(a^{\downarrow_1},b^{\downarrow_1})$ (in \inc{\spmn}{1}).
We will show that $d = c^{\downarrow_1}$.
We have $c\ge a$ and $c \ge b$, therefore
$c^{\downarrow_1} \ge a^{\downarrow_1}$ and $c^{\downarrow_1} \ge
b^{\downarrow_1}$, which implies that $c^{\downarrow_1} > d$.
Let us begin by showing that $d_1 = c_1 + 1$. We can suppose $a_1 \ge b_1$.
The partition $(a_1, a_1, a_1-1, a_1-2, \dots )$ is greater than $a$ and $b$,
hence it is greater than $c$. This implies that $c_1 = a_1$. Since $\inc{a}{1}
\le d \le \inc{c}{1}$, we have $d_1 = a_1 + 1$. Let $e=(d_1-1, d_2, d_3,
\dots)$. Since $d \le \inc{c}{1}$, $e$ verifies the characterisation of
Theorem \ref{CaractElts}. Moreover, $d \ge \inc{a}{1}$ and $d \ge \inc{b}{1}$,
hence $e
\ge a$ and $e \ge b$. This implies that $e \ge \sup(a,b) = c$ and that
$d \ge \inc{c}{1}$, and so $d = \inc{c}{1}$.
%
\cqfd

\fig{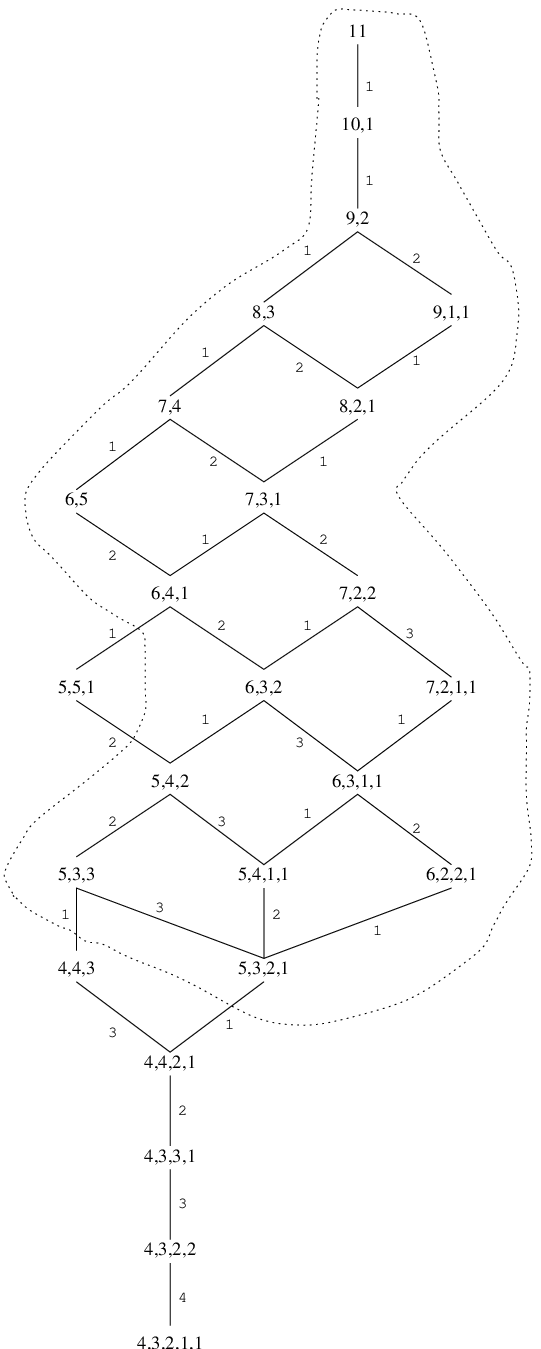}{$SPM(10)^{\downarrow_1}$ in $SPM(11)$.}

\noindent
It is straightforward that each element $s$ of \spmnp\ is reachable from an element of \spmn\fb{1}.
Indeed, $s$ is at least reachable from $\inc{(n)}{1} = (n+1)$. This shows that
one can start the construction of \spmnp\ with \spmn\fb{1}\ and then add
the missing elements (see Figure \ref{fig_spm10_11} for an example).

The construction procedure starts with
the lattice \spmn\fb{1}\ given by its diagram. Then, we look for those
elements in \spmn\fb{1}\ that have a successor out of \spmn\fb{1}. The set of
these elements will be denoted by \I{1}, with $\I{1} \subseteq \spmn\fb{1}$.
At this point, we add all the missing successors of the elements of $I_1$.
The set of these new
elements will be denoted by \C{1}. Now, we
look for the elements in \C{1}
that have a successor out of the constructed set. The set of these elements is
denoted by \I{2}. We add the new elements (their set is denoted by \C{2}),
and we iterate this process until the set $I_i$ is empty.

More explicitly, in the $i$-th step of the procedure we
look for the elements in \C{i-1} with missing
successors and call \I{i} the set of these elements. We add the new
successors of the elements of \I{i} and call the set of these new elements
\C{i}. At each step, when we add a new element, we also add its covering
relations. \spmnp\ is a finite set, therefore this procedure terminates.
At the end, we have
obtained the whole set \spmnp\ with its order relation.

Now, let us show how this completion of \spmn\fb{1}\ to obtain \spmnp\ is
implemented.
Recall that any element $t$ of \spmn\fb{1}\ is
obtained from an element
$s$ of \spmn\ by adding a new grain on the first column.
Three cases are possible:
\begin{itemize}
\item $s$ begins with a plateau or a cliff. Then, according to Proposition
      \ref{Prop_plateau}, the possible transitions from \inc{s}{1}\ 
      are the same as the possible transitions from $s$,
      and the successors of \inc{s}{1}\ 
      are obtained by an application of \fb{1}\ to the successors of $s$.
      Therefore they are already in \spmn\fb{1}\ (see Figure
      \ref{fig_nouv_elts}(a)(b)).
\item $s$ begins with a step. In this case, $s$ is in
      $P_1 = \spmn\di{1}{1}$. Then, according to Proposition
      \ref{Prop_escalier}, only one successor of \inc{s}{1}\ is not yet in
      \inc{\spmn}{1}: the successor obtained by a transition \tr{1}.
      This element is \inc{s}{2}\ (see Figure \ref{fig_nouv_elts}(c)).
      It follows that $\I{1}=\inc{P_1}{1}$ and $\C{1}=\inc{P_1}{2}$.
\end{itemize}

\fig{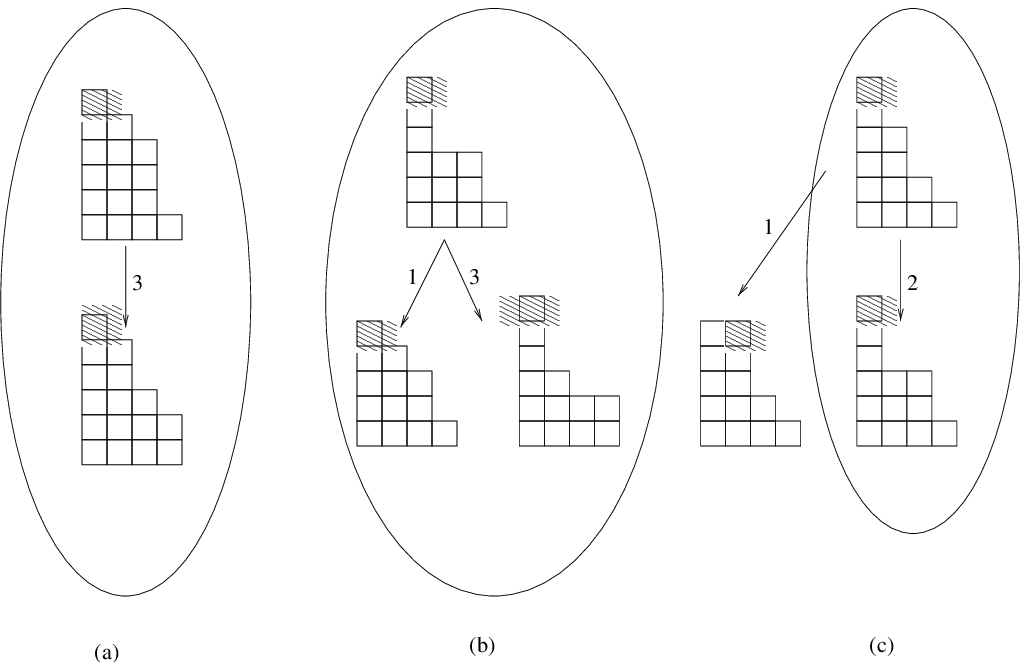}{ \
Successors of \inc{s}{1}\ if $s$ begins with (a) a plateau \
(b) a cliff and (c) a step: there is a new element only in this last case.}

This means that the first step of the construction consists in adding all the
elements of $\C{1} = \inc{P_1}{2}$. Notice that this set is
added with a duplication of
the order structure of $\inc{P_1}{1} = \I{1}$. Indeed, it is clear that:
$$
(\inc{s}{1} \tr{j} \inc{t}{1})\mbox{ iff }(\inc{s}{2} \tr{j} \inc{t}{2})
\mbox{ for all $s$, $t$ in $P_1$, and for all $j$.}
$$

The following step consists in adding the missing successors
of the elements of \C{1}
and the missing transitions originating from them. The analysis of the three
cases (plateau, cliff, step) shows that the only elements of \C{1} that
do not have all their successors and transitions are:
\begin{itemize}
\item The elements $s^{\downarrow_{2}} \in \C{1}$ such that
      $s$ has a cliff at $2$. Indeed, such a $s^{\downarrow_2}$ does have all its
      successors in the lattice, but one transition is missing:
      the one labelled
      with $2$. In this case Proposition \ref{Prop_falaise} shows that
      $s^{\downarrow_{2}} \stackrel{2}{\longrightarrow} t$ where
      $t$ is also obtained by
      $s^{\downarrow_{1}} \stackrel{2\cdot 1}{\longrightarrow} t$.
      Therefore, we have to add an edge $\stackrel{2}{\longrightarrow}$ from
      $s^{\downarrow_{2}}$ to an element $t$
      which is already in the lattice. We will
      call \emph{back edge} such an edge.
\item The elements $s^{\downarrow_2} \in \C{1}$ such that
      $s$ has a step at $2$ (i.e. $s$ begins with stairs of length at
      least $2$ and hence is in $P_2^{\downarrow_2}$). According to
      Proposition \ref{Prop_escalier}, only one successor of each of these
      elements is not yet in the lattice: the successor obtained by
      the transition on the second column, i.e. the element \inc{s}{3}.
      Therefore, to complete the second step, we have to add the set
      \inc{P_2}{3} (with the same order structure
      as \inc{P_2}{2}) to the existing lattice
      and connect the lattice to this new part by all the transitions:
      $$ s^{\downarrow_{2}} \stackrel{2}{\longrightarrow}
      s^{\downarrow_{3}}
      \hspace{0.5cm}\mbox{ for all }s^{\downarrow_{2}} \in P_2^{\downarrow_{2}}.$$
\end{itemize}

This means that $\I{2} = \inc{P_2}{2}$ and $\C{2} = \inc{P_2}{3}$.
In general, the $i$-th step consists in adding the missing successors of the
elements added at step $i-1$ and the missing transitions originating from them.
We show that the observed behaviour for the second step is general,
and so the sets \I{i} and \C{i} can be characterized.

\begin{theo}
For all integer $i$, we have 
$\I{i} = \inc{P_i}{i}$ and $\C{i} = \inc{P_i}{i+1}$.
\label{}
\end{theo}
\proof
By induction:
\begin{itemize}
\item The case $i=1$ has already been studied. Notice that every
 covering relation
 concerning the new elements is of the following form: \inc{s}{1} \tr{1}
 \inc{s}{2}\ where $s \in P_1$.
\item Suppose the result is true for $i-1$. We show that it is
 true for $i$. Consider $\C{i-1} = \inc{P_{i-1}}{i}$. Using Propositions
 \ref{Prop_plateau}, \ref{Prop_falaise} and \ref{Prop_escalier}, we look for
 the successors of \inc{s}{i}, with $s \in P_{i-1}$. Three cases are to be
 considered: $s$ can have a plateau, a cliff or a step at $i$.
 \begin{description}
 \item[Plateau] According to Proposition \ref{Prop_plateau},
   $Succ(\inc{s}{i}) \subseteq \inc{P_{i-1}}{i} = \C{i-1}$. So,
   $s$ has no new successor, and $s \not\in \I{i}$. Moreover,
   $s \tr{j} t$ if and only if $\inc{s}{i} \tr{j} \inc{t}{i}$.
 \item[Cliff] According to Proposition \ref{Prop_falaise},
   $Succ(\inc{s}{i}) \subseteq \inc{P_{i-1}}{i} \cup \inc{\spmn}{1}$.
   So, $s \not\in \I{i}$. Moreover, the edges of the covering relation
   originating from \inc{s}{i}\ are the same
   than the ones originating from $s$ plus an additional one:
   \inc{s}{i} \tr{i} \inc{s}{i+1}. The element \inc{s}{i+1}\ is in
   \inc{\spmn}{1}\ according to Proposition \ref{Prop_falaise}.
 \item[Stair] According to Proposition \ref{Prop_escalier},
   \inc{s}{i}\ has a new successor, \inc{s}{i+1}, hence $\inc{s}{i} \in \I{i}$.
   Moreover, the edges of the covering relation originating from \inc{s}{i}\ are
   the same as   the ones originating from $s$ plus an additional one:
   \inc{s}{i} \tr{i} \inc{s}{i+1}. Therefore the element \inc{s}{i+1}\ is in \C{i}.
 \end{description}
 From these three cases, we deduce the claim.
 \cqfd
\end{itemize}

We have obtained a characterization of the sets \I{i} and \C{i}.
It is now straightforward that Algorithm \ref{Algo_incr}
constructs the lattice \spmnp\ from \spmn. Notice that we can obtain \spmn\ 
for an arbitrary integer $n$ by starting from $SPM(0)$ and iterating this algorithm.
In the next sections, we will give more details about this construction. We
will show that the complexity of Algorithm \ref{Algo_incr} is linear with
respect to
the number of newly added elements, and hence we have an algorithm that constructs
\spmn\ in linear time linear with respect to $|\spmn |$.

\begin{algorithm}
\SetVline
\In{\spmn}
\Out{\spmnp}
\Begin{
$i \leftarrow 1$\;
$I \leftarrow \inc{P_i}{i}$\;
\While{$I \neq \emptyset$}{
  $C \leftarrow \inc{P_i}{i+1}$\;
  add $C$ with its covering relation\;
  \ForEach{$\inc{s}{i}\ in\ I$}{add the edge:
         $\inc{s}{i} \tr{i} \inc{s}{i+1}\in C$}
  \ForEach{$\inc{s}{i+1}\ in\ C$ s.t. $d_{i+1}(s)\ge 2$}{add the back edge
         $\inc{s}{i+1} \tr{i+1} s' \in \inc{\spmn}{1}$}
  $i \leftarrow i+1$\;
  $I \leftarrow \inc{P_i}{i}$\;
}}
\caption{Incremental construction \label{Algo_incr}}
\end{algorithm}

\subsection{Structure of the $P_i$ parts}

We will now study more deeply the construction procedure given above.
We will obtain results on the structure of the $P_i$ parts, which play
an important role, and a recursive formula for $|\spmn|$.
However, the results presented here are not necessary to understand the
infinite extension presented in the second part of the paper.
Therefore, the rest of this section can be ignored if the reader
is mostly interested
in the second part of the paper.

In the previous section, we characterized the sets
\I{i} and \C{i}. More can be said about the
structure of these sets. In fact, since $\I{i} = \inc{P_i}{i}$ and $\C{i} =
\inc{P_i}{i+1}$, we only have to study the sets $P_i$. We will show
that these sets are disjoint unions of lattices,
and that each of these lattices is obtained
from a generating partition by iteration of the \spm\ rule. We will give the
explicit characterisation of these generating partitions,
as well as their number.

\begin{prop}
\label{prop_gene}
$P_1$ is a disjoint union of lattices.
\end{prop}
\proof
Let $Q_{1,k}$ denote the set of all the elements
of \spmn\ whose first two parts are $k$ and $k-1$.
This is a non-empty subset of $P_1$.
It is clear that if $k \neq k'$ then $Q_{1,k} \cap Q_{1,k'} = \emptyset$,
so $P_1$ is the disjoint union of the sets $Q_{1,k}$.

Since \mbox{$P_1 \subseteq
\spmn$}, the elements of $Q_{1,k}$ verify the characterisation of
Theorem \ref{CaractElts}; this implies that the maximal element $g$ of $Q_{1,k}$
has the form
$$g = (k,k-1,k-1,k-2,\ldots,k-l,r)$$
with $l$ maximal (i.e. $l$ such that $r\leq k-l-1$),
and $k+(k-1)+(k-1)+(k-2)+\ldots+(k-l)+r=n$.
Then $g$ belongs to
\spmn\ if and only if $k$ satisfies:
$$k+k-1 \leq n \leq k+(k-1)+(k-1)+(k-2)+\ldots+1$$
which is equivalent to $2k-1 \leq n \leq \frac{k^2+3k-2}{2}$.

Let $k$ be such an integer.
Let us study the structure of $Q_{1,k}$ by considering its maximal
element $g$, described as above. Let $s$ be an element of $Q_{1,k}$. It is
clear that the prefix sums of $s$ are less than or equal to the
ones of $g$, so, according
to \cite{HDP3}, $s$ can be obtained from $g$ by the \spm\ rule.
Therefore, $Q_{1,k}$ is the set of the elements of \spmn\ which can be
obtained from $g$ and whose first two parts are $k$ and $k-1$.
In other words, $Q_{1,k}$ is the set of the
partitions reached from $g$ by paths without any transition labelled
$1$ or $2$. The element $g$ is called the \emph{generating partition}
of $Q_{1,k}$.
Therefore $Q_{1,k}$ is isomorphic to the lattice of the partitions of $n-2k-1$
obtained from $(g_3,\ldots,g_n)$ by iteration of the \spm\ rule, and,
in particular, $Q_{1,k}$ is a lattice.
\cqfd

\noindent
More generally, let us denote by $Q_{i,k}$ the set
$Q_{1,k} \cap P_i$.

\begin{prop}
The sets $Q_{i,k}$ are lattices with all transitions labelled
with integers
greater than $k$. Moreover, for all $i$, $P_i$ is the disjoint union of the
lattices $Q_{i,k}$.
\end{prop}
\proof
Recall that $P_i$ is the subset of $P_1$ containing the partitions that begin
with stairs of length $i$, and that $Q_{1,k}$ is the subset of $P_i$
containing the partitions that begin with $k,k-1$. So, $Q_{i,k}$ is the
subset of $P_1$ containing the partitions that begin with the stairs
$k,k-1,\ldots,k-i$. Therefore the maximal element of $Q_{i,k}$ has the form:
$$g=(k,k-1,\ldots,k-i,k-i,\ldots,k-l,r)$$
where $l\geq i$ and $l$ is maximal (i.e. $r \leq k-l-1$) (we use here the
same argument as in the proof of
Proposition \ref{prop_gene}).
Every element $s$ of $Q_{i,k}$ is reachable
from this element $g$,
and only transitions with labels greater than $i$ are needed to obtain $s$
from $g$. Therefore, $Q_{i,k}$
is a lattice isomorphic to the lattice of partitions obtained from
$(k-i,\ldots,k-l,r)$.
We have $$P_1 = \bigsqcup_k Q_{1,k}$$ where $\sqcup$ denotes the disjoint
union, and the $Q_{1,k}$ are pairwise
disjoint, then:
$$
P_i = P_1 \cap P_i = \bigsqcup_k(Q_{1,k}\cap P_i) = \bigsqcup_k Q_{i,k},
$$
and obviously the sets $Q_{i,k}$ are also pairwise disjoint for a fixed $i$.
\cqfd

\noindent
We sum up these results in Figure \ref{fig_recap}.
An example is given for $n=10$ in Figure \ref{fig_recap10_11}.

\fig{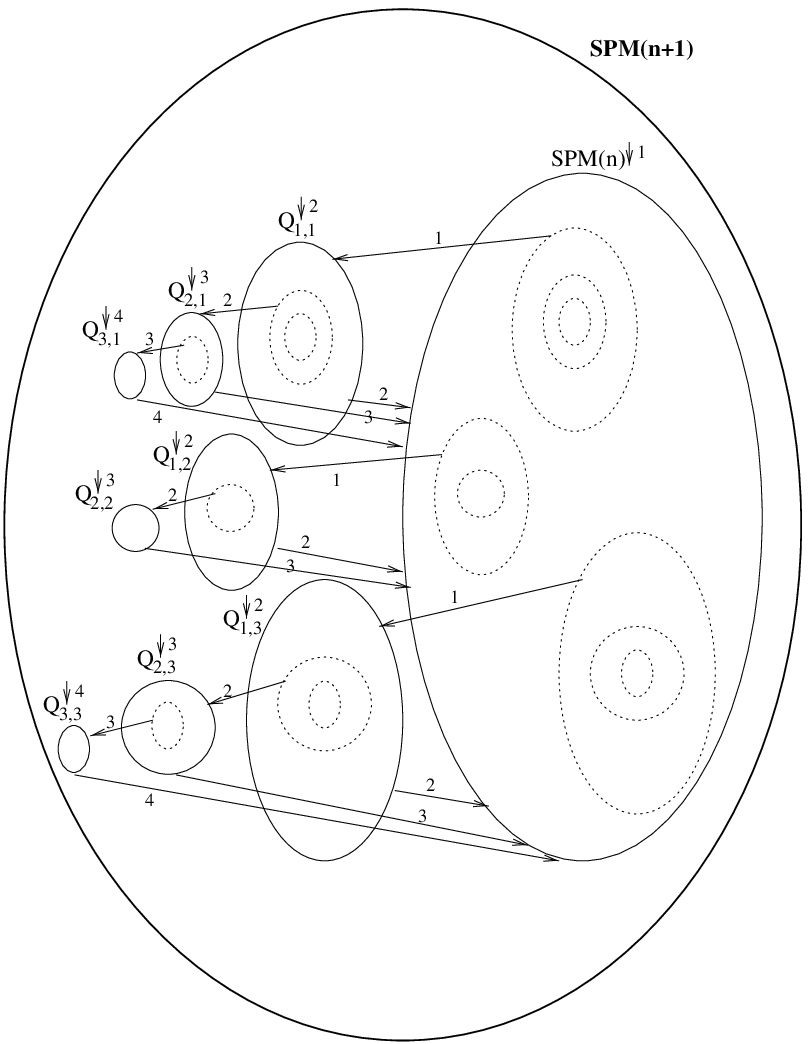}{Structure of SPM(n+1).}

\fig{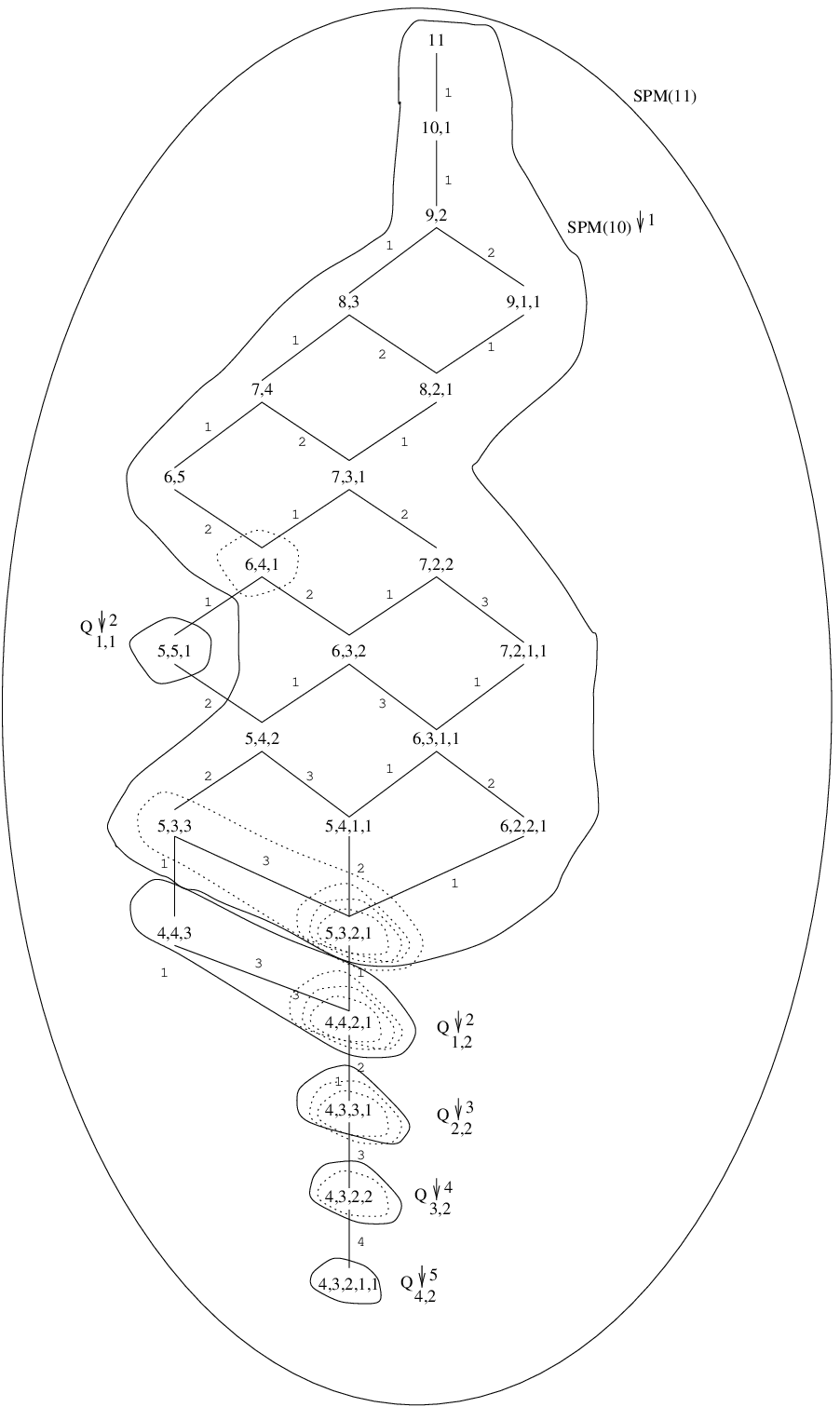}{Example with n=10.}

We have defined the generating partition $g$
of a set $Q_{1,k}$ as the maximal element of $Q_{1,k}$. Therefore
$Q_{1,k}$ is the lattice of the sand piles reachable from $g$ by iteration of
the \spm\ rule on the coulmns at the right of the second column.
Using the characterisation of \spm\ partitions, we can now enumerate all the
generating partitions for a given $n$.

\begin{prop}[Generating partitions]
The number of generating partitions in the set \spmn\ is
$\lfloor \frac{n}{2}+2-\sqrt{\frac{17}{4}2n} \rfloor$.
\end{prop}
\proof
As seen above, the generating partitions in \spmn\ have the form:
$(k,k-1,k-1,k-2,\dots ,k-l,r)$ for some $l > 0$ and $r \le k-l-1$.
Such integers $k$ must verify
$k + (k-1) \le n$ and $k-1 + \frac{k(k+1)}{2} \ge n$.
Moreover, any of these $k$ effectively corresponds to a generating partition.
Therefore we have as many generating partitions as solutions to the system:
$$
\left\lbrace
\begin{array}{l}
k-1 + \frac{k(k+1)}{2} \ge n \\
k + (k-1) \le n,
\end{array}
\right.
$$
that is
$ -\frac{3}{2} + \sqrt{\frac{17}{4} + 2n} \le k \le \frac{n+1}{2} $.
\cqfd

We have already seen that each $Q_{1,j}$ is a lattice and contains
the lattices $Q_{i,j}$ for $i > 1$.
The lattices $Q_{i,j}$ also verify: $Q_{i+1,j} \subseteq Q_{i,j}$
for all $j$. We will show how the $Q_{1,j}$ are generated during the
construction. See Figure \ref{fig_copies}.

\fig{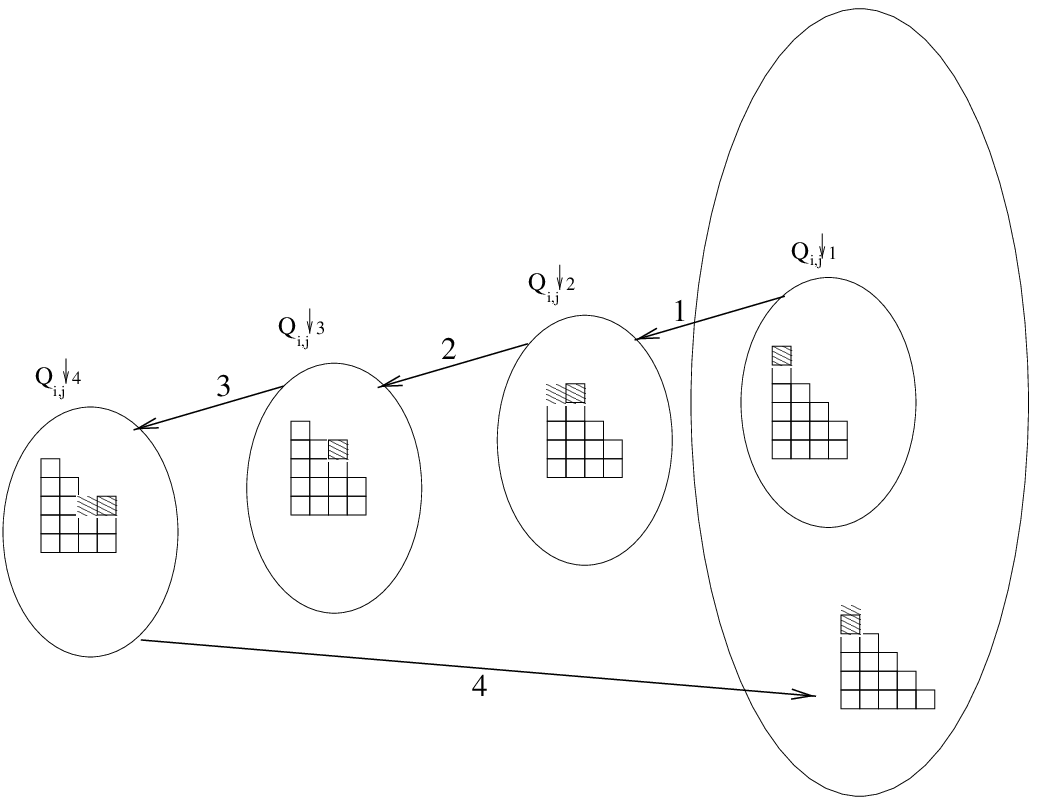}{Successive copies of a $Q_{i,j}$.}

In order to study the parts $P_i$ and $Q_{i,j}$ when $n$ varies,
let us extend our notations.
We will denote by $P_i(n)$ the parts $P_i$ of $SPM(n)$ ($P_i(n)$ is the set
of all sand piles
with $n$ grains that begins with stairs of length at least $i$).
Likewise, $Q_{i,j}(n)$ denotes the part $Q_{i,j}$ of $SPM(n)$.
We can make three remarks:
\begin{itemize}
\item The elements obtained from $P_i(n)$ by applying the operator
  \fb{1}\ any number of times do not belong to a $P_i(m)$ with $m>n$
  since they begin with a cliff.
\item The elements of \inc{(P_i(n))}{2}\ begin with a plateau at column $1$
  followed by a cliff at column $2$. Then, if we apply \fb{1}\ to these
  elements we obtain sand piles which
  begin with stairs of length exactly $1$ (i.e.
  $(P_i(n))^{\downarrow_2\downarrow_1} \subseteq P_1(n+2))$.
  Likewise,
  $(P_i(n))^{\downarrow_k \downarrow_{k-1} \dots \downarrow_2 \downarrow_1}
  \subseteq P_{k-1}(n+k)$.
\item The elements of \inc{(P_i(n))}{k}\ $(k \le i)$ begin with a stair
  of length exactly $k-1$ and hence are in $P_k(n+1)$.
\end{itemize}

\noindent
From these remarks we deduce the result:
\begin{theo} Let $i$ be an integer such that
$\frac{i(i+1)}{2} \le n+2$. Then:
\label{Theo}
\begin{equation}
\label{Equ_unions}
P_i(n+2) = T(n+2) \sqcup
 (P_i(n-i+1))^{\downarrow_1\downarrow_2\dots \downarrow_i\downarrow_{i+1}}
 \sqcup \left( \bigsqcup_{k>i}(P_k(n+1))^{\downarrow_{k+1}} \right)
\end{equation}
where $T(n+2)$ is a set that contains one partition at most, namely:
$$
T(n+2) =
\left\lbrace
\begin{array}{ll}
\lbrace (k,k-1,\dots ,2,1) \rbrace & if\ \exists\ k\ integer\ s.t.\ n+2=\frac{k(k+1)}{2}\\
\emptyset & otherwise
\end{array}
\right.
$$
with the initial conditions:
$$
\left\lbrace
\begin{array}{l}
P_1(1) = \lbrace (1,0,\dots) \rbrace\\
P_i(1) = \emptyset\ \forall i > 1\\
P_i(2) = \emptyset\ \forall i.
\end{array}
\right.
$$
\end{theo}
\proof
$P_i(n+2)$ contains each of the sets in the right hand side of Equation
\ref{Equ_unions}. Indeed,
if we add one grain on each of the $i+1$ first columns of a sand pile with $n-i+1$
grains that begins with a stair of length $i$, we obtain a sand pile with
$n$ grains which also begins with a stair of length $i$.
If we take a sand pile of $n+1$ grains which
begins with a stair of length at least
$i$, and we add a grain on column $k+1$ with $k>i$, we obtain a sand pile of
$n+2$ grains which also begins with a stair of length at least $i$. Finally,
if $n+2$ has the form $\frac{k(k+1)}{2}$ for
some integer $k$ and if $i$ is
smaller than or equal to $k$ (the length of the sand pile $(k,k-1,\dots,1)$)
then this element of $T(n+2)$ begins with a stairs of length at least $i$.

Likewise, each element of $P_i(n+2)$ is in one of those sets.
Let $s$ be in $P_i(n+2)$. Three cases are possible:
 \begin{itemize}
  \item $s$ has a step at each column, i.e. $s \in T(n+2)$.
  \item $s$ begins with stairs of length $k$ with $k \geq i$ and $s$ has a
     plateau at $k+1$. Then, it is an element of
     $(P_{k+1}(n+1))^{\downarrow_{k+2}}$. We know that such elements exist
     from the characterisation of Theorem
     \ref{CaractElts}.
  \item $s$ begins with a stair of length $k$ with $k \ge i$ and $s$ has a
     cliff at $k+1$. Then, $s$ is an element of
     $(P_i(n-i+1))^{\downarrow_1\downarrow_2\dots\downarrow_{i+1}}$.
     We know that such elements exist from the characterisation of
     Theorem \ref{CaractElts}.
 \end{itemize}

Now, let us show that the unions in Formula \ref{Equ_unions} are disjoints.
The elements of the set \mbox{\inc{(P_k(n+1))}{k+1}} with $k>i$ begin
with stairs of length exactly $k$. So, the set \inc{(P_k(n+1))}{k+1} and
\inc{(P_{k'}(n+1))}{k'+1} with $k,k'>i$ are
pairwise disjoints. Moreover, the set
\mbox{\inc{(P_i(n-i+1))}{1}\fb{2}\fb{\dots}\fb{i}\fb{i+1}} only contains
elements that begin with stairs of length exactly $i$, so they doesn't
intersect the parts \inc{(P_k(n+1))}{k+1}\ which begin with stairs of length
$k$ with $k>i$. 
Finally, if $T(n+2)$ is non-empty, its element clearly does not belong to any
of the other sets.
\cqfd

This theorem gives a better understanding of the structure of
the lattices \spmn. Since the unions are disjoints,
the formula is even more interesting as it gives a way
to compute the cardinality of \spmn. We first state the
following corollary, immediate from Theorem \ref{Theo}.

\begin{corol}
\label{CorolCardinal}
Let $p_{i,n}$ denote  $| P_i(n) |$, i.e.
$p_{i,n}$ is the number of partitions in \spmn\ that begin with stairs of
length at least $i$. We have:
$$
p_{i,n+2} = p_{i,n-i+1} + \sum_{j>i} p_{j,n+1} + \delta_{i,n+2}
$$
where
$$
\delta_{i,n+2} = 
\left\lbrace
\begin{array}{ll}
 1 & \mbox{if }\exists k\ \in\ \mathbb{N}\ s.t.\ n =
\frac{k(k+1)}{2}\\
 0 & otherwise
\end{array}
\right.
$$
with the initial conditions
$p_{1,1} = 1,\ p_{1,j} = 0$ for all $j>1$, $p_{j,2} = 0$ for all $j > 0$, and $p_{0,2}=2$.
\end{corol}

\noindent
This corollary gives a way to compute the number of elements in \spmn\ since
$|SPM(1)| = 1$ and
$$
| SPM(n) | = |SPM(n-1)| + \sum_i p_{i,n-1}.
$$
Notice that this formula is nothing but the formula of the Corollary
\ref{CorolCardinal} specified for $i = 0$. This is not surprising, since any element of
\spmn\ begins with stairs of length at least $0$.


\section{Infinite extension of SPM}

Let us now present an infinite extension of \spm. Two different
possible generalisations are natural to extend the notions
studied until here. The first one is to consider a column
with an infinite number of grains as the initial configuration,
and then study the evolution of the system with respect to the
\spm\ rule. We call this model \spmi. The second one is to use
the construction detailed in the first part of the paper to
extend the order on $\bigsqcup_{n\ge 0} \spmn$. It turns out
that these two ideas lead to two isomorphic objects. This
gives us an efficient way to construct \spmi, as shown below.
Afterwards, we introduce the infinite tree \spt, and we show
a possible coding of \spmi\ using this tree. The study of the
properties of this tree gives a new recursive formula to compute
$|\spmn|$.

\subsection{The infinite lattice \spmi}
\spmn\ is the lattice of the configurations reachable from
the partition $(n)$ by iteration of the \spm\ rule. We will
now define \spmi\ as the set of all configurations reachable
from $(\infty)$ (this is the configuration where the first
column contains infinitely many grains).
The covering relation on \spmi\ is defined by: $s \tr{i} t$
if and only if $t$ is obtained from $s$ by application of
the \spm\ rule on the $i$-th column. The order on \spmi\ is
the reflexive and transitive closure of this covering relation.
Notice that any element $s$ of \spmi\ has the form
$(\infty,s_2,s_3,\dots,s_k)$. The first partitions in
\spmi\ are given in Figure \ref{fig_spm_infi} along with
their covering relations (the first column, which always
contains an infinite number of grains, is not represented
on this diagram).

\fig{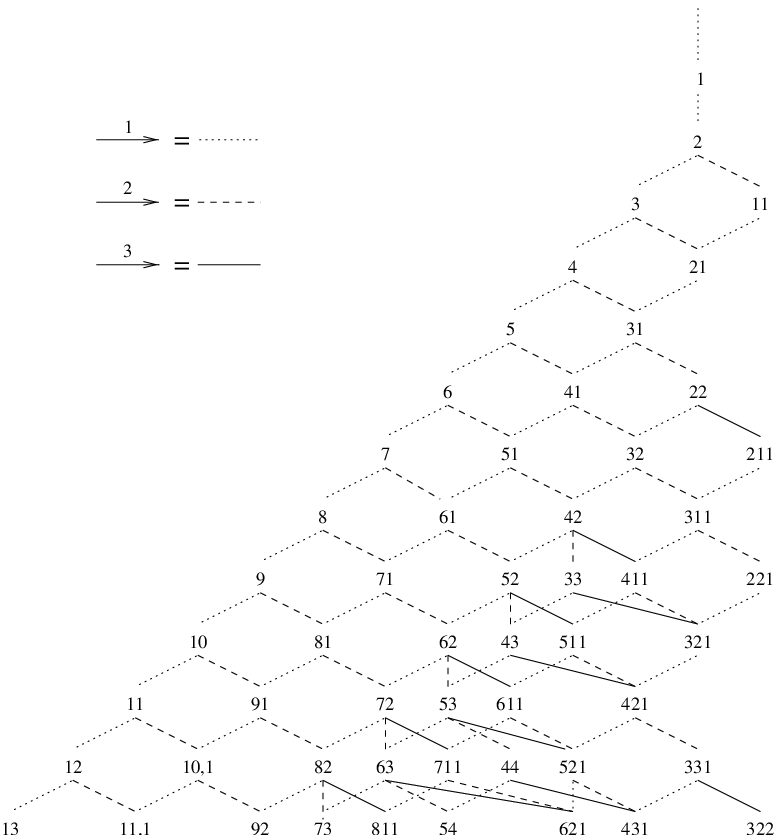}{First elements and transitions of \spmi.}

Notice also that the first column does not influence the
characterisation of the elements given in Theorem \ref{CaractElts}.
We will now show that \spmi\ is a lattice.
To do so, we will use the notion of \emph{shot vector}
(see \cite{TheseEriksson}, section 5.3). The shot vector $k(s,t)$
from the sand pile $s \in \spmn$ to the sand pile $t \in \spmn$
is defined by the following:
the $i$-th component $k_i(s,t)$ of $k(s,t)$ is the number
of applications of the \spm\ rule on column $i$ in order
to obtain $t$ from $s$.

We need here an extension of this definition:
the $i$-th component of the shot vector $k((\infty),s)$
from $(\infty)$ to $s \in \spmi$ is the number of applications
of the \spm\ rule on column $i$ in order to obtain $s$ from
$(\infty)$. It is straightforward to see that $k((\infty),s)$
is given by:
$$
k_1((\infty),s) = s_2 + s_3 + \dots
$$
$$
k_2((\infty),s) = s_3 + s_4 + \dots
$$
and in general:
$$
k_i((\infty),s) = \sum_{j>i} s_j
$$

\noindent
From \cite{HDP3}, we have:
\begin{lemme}
\label{lemme_shot}
Let $s$ and $t$ be two elements of \spmi. Then,
$$
s \ge t\ \mbox{ iff }\ k((\infty),s) \le k((\infty),t).
$$
Moreover, if $m$ denotes the max of $k((\infty),s)$ and $k((\infty),t)$ then the
partition $u$ such that $k((\infty),u) = m$ is in \spmi\ and $u = \inf(s,t)$.
\end{lemme}

\noindent
With this result, we can show that \spmi\ is a lattice:
\begin{theo}
The set \spmi\ is a lattice. Moreover, let $s=(\infty,s_2,\dots,s_k)$ and
$t=(\infty,t_2,\dots, t_l)$ be two elements of \spmi, then, $\inf(s,t)=u$
in \spmi, where
$$
u_i = max(\sum_{j\geq i} s_j, \sum_{j \geq i} t_j) - \sum_{j >i} u_j
\hskip 0.3cm \mbox{ for all } i\mbox{ such that } 2 \leq i\leq max(k,l)
$$
and $\sup(s,t) = \inf \lbrace u \in SPM(\infty), u \geq s, u \geq t \rbrace $.
\end{theo}
\proof
From Lemma \ref{lemme_shot} and the definition of the shot vectors in \spmi,
we have the formula for the infimum. Since $(\infty)$ is the maximal element of \spmi,
this set is a lattice. \cqfd

From the definition, it is possible to show that \spmi\ contains an isomorphic copy of \spmn\ for any integer $n$.
\begin{prop}
Let $n$ be a positive integer. The application:
$$
\begin{array}{lclc}
\pi : & \spmn                   & \longrightarrow & \spmi\\
      & s=(s_1, s_2,\dotsi,s_k) & \longrightarrow & \bar{s} = (\infty,s_2,\dots,s_k)
\end{array}
$$
is a lattice embedding, which means that it is injective
and preserves the infimum and the supremum.
\label{prop_pi}
\end{prop}
\proof
Again, we will use the shot vector $k(a,b)$ from $a$ to $b$.
Recall that $k_i(a,b)$ is nothing but the number of grains falling
from column $i$ in order to obtain $b$ from $a$.
Let $s$ and $t$ be in \spmn. Suppose $\overline{s} = \overline{t}$, i.e.
$s_i = t_i$ for all $i\ge 2$. We have $$s_1 = n - \sum_{i\ge 2} s_i = n - \sum_{i\ge 2}
t_i = t_1,$$ hence $s=t$ and $\pi$ is injective.

It is clear that $a \tr{i} b$ in \spmn\ if and only if $\bar{a} \tr{i}
\bar{b}$ in \spmi, hence $a \geq b$ in \spmn\ if and only if $\bar{a}
\geq \bar{b}$ in \spmi. So, in this case, $k(a,b) = k(\bar{a}, \bar{b})$.
Let $c$ be the infimum of two elements $a$ and $b$ in \spmn.
We show that $\bar{c} = \inf(\bar{a},\bar{b})$ in \spmi. Since
$c=\inf(a,b)$ in \spmn, we have: $$k((n),c) =max (k((n),a),k((n),b)).$$ Moreover,
$k((n),a))=k((\infty),\bar{a})$ for all element $a$ of \spmn, hence we can deduce
\mbox{$k((\infty),\bar{c}) =max(k((\infty),\bar{a}),k((\infty),\bar{b})$}, and
$\bar{c} = \inf(\bar{a},\bar{b})$ in \spmi, as expected. So, the infimum
is preserved.

Now, let us prove that the supremum is preserved.
Let $d=(d_1,\ldots,d_n) =\sup(a,b)$ in \spmn, and let $e=(\infty,e_2,\ldots,e_m)
= \sup(\bar{a},\bar{b})$ in \spmi. We must show that $\bar{d} =e$. To do so, we
show that an element $f$ of \spmn\ such that $e=\bar{f}$ exists.
Since $d\geq a$ and $d \geq b$ in \spmn, we have $\bar{d} \geq
\bar{a}$ and $\bar{d} \geq \bar{b}$ in \spmi, hence $\bar{d} \geq e$ in \spmi,
and $k((\infty),\bar{d})\leq k((\infty),e)$. We first show that 
$k_1((\infty),\bar{d})= k_1((\infty),e)$. Without loss of generality, we can
suppose that $a_1 \geq b_1$. Notice that the partition
$(a_1,a_1,a_1-1,a_1-2,\ldots)$ is greater than $a$ and $b$, hence
greater than $d$, and so $a_1 \geq d_1$. Since $d \geq a$, we have $d_1 = a_1$
and
$k_1((\infty),\bar{d}) =k_1((\infty),\bar{a})$. Moreover, $\bar{a} \leq e \leq
\bar{d}$, hence $k_1((\infty),\bar{d)}= k_1((\infty),e)$. Let us define
$f=(n-e_2-\ldots-e_m,e_2,\ldots,e_m)$. Since $e\leq \bar{d}$, we have $e_2
\leq d_2$ and since $e$ verifies the characterisation of Theorem \ref{CaractElts},
so does $f$, hence $f \in SPM(n)$ and $e=\bar{f}$. Since
$e \geq \bar{a}$ and $e \geq \bar{b}$, we have $f \geq a$ and $f\geq b$, hence
$f\geq d$. This implies that $e=\bar{f} \geq \bar{d}$. Therefore $e=\bar{d}$.
This gives the result.

\cqfd

Let $\overline{\spmn}$ denotes the image by $\pi$ of \spmn\ in \spmi.
From Proposition \ref{prop_pi}, $\overline{\spmn}$ is a sublattice
of \spmi. By Theorem \ref{prop_sub}, \inc{\spmn}{1} is a sublattice
of \spmnp, hence, since $\overline{\inc{\spmn}{1}} = \overline{\spmn}$,
we have an increasing sequence of sublattices:
$$
\overline{SPM(0)} \leq \overline{SPM(1)}\leq
 \dots \leq \overline{SPM(n)} \leq \overline{SPM(n+1)} \leq \dots \leq \spmi,
$$
where $\le$ denotes the sublattice relation.

Let $s=(\infty,s_2,s_3,\dots,s_k)$ be an element of \spmi, then $s$ verifies the characterisation of Theorem \ref{CaractElts}. If one takes $s_1 = s_2 + 1$ and $n = \sum_{i=1}^{k} s_i$, we have that $s' = (s_1,s_2,\dots,s_k)$ is an element of \spmn. This implies that $s=\pi(s')$ and that $s$ is an element of $\overline{\spmn}$, therefore:
$$
\bigcup_{n\ge 0} \overline{\spmn} = \spmi
$$


Let us now study the disjoint union of the lattices \spmn\ for $n \ge 0$.
Let us define
$$ S = \bigsqcup_{n\ge0}\spmn,$$
on which we extend the order relation of each \spmn\ as follows.
Let $s \in SPM(m)$ and $t \in SPM(n)$. We define
$s \tr{i} t$ in $S$ if and only if we are in one of
the two following cases:
\begin{itemize}
\item $n=m$ and $s \tr{i} t$ in \spmn
\item $i=0$, $n = m+1$ and $b=\inc{a}{1}$.
\end{itemize}
In other terms, the elements of \spmn\ are linked to each other as usual
whereas each element $a$ of \spmn\ is linked to $\inc{a}{1} \in \spmn$ by an
edge labelled $0$.
From this covering relation, one can define an order on the set $S$
as the reflexive and transitive closure of this covering relation.

\begin{theo}
For all integer $n$, \spmn\ is a sublattice of $S$.
\end{theo}
\proof
The fact that \spmn\ is present in $S$ is immediate from the definition.
What we have to show is that the lattice structure of \spmn\ is
preserved in $S$.
Let $s$ and $t$ be two elements of \spmn. We have to show that the infimum and the
supremum of $s$ and $t$ in $S$ are in \spmn.
Let $u=\inf(s,t)$ in \spmn\ and $u'=\inf(s,t)$ in $S$. We have that
$s \geq u' \geq u$ in $S$, thus 
$$\sum_{i\geq 1} s_i \leq \sum_{i\geq
1}u'_i  \leq \sum_{i\geq 1}u_i,$$
and $$\sum_{i\geq 1} u'_i =n.$$
Therefore $u'$ is an
element of \spmn, and we have $u'=u$. The same method can be applied for the supremum.
\cqfd

\noindent
This result is illustrated in Figure \ref{fig_SPM_S} (right).
The surprising result is that these two ways to extend the sand pile model
to infinity, i.e. the first one by adding new elements to some \spmn\ to
extend it into \spmnp\ and infinitely iterating the process to obtain
\spmi, and the second one by linking together all the \spmn\ for all $n$
to obtain $S$, lead to the same object.
\begin{theo}
\label{th_chi}
The application $\chi$ defined by:
$$
\chi : S = \bigsqcup_{n\geq 0}SPM(n) \longrightarrow SPM(\infty)
$$
$$
s=(s_1,s_2,\ldots,s_k) \mapsto \chi(s) = (\infty,s_1,s_2,\ldots,s_k)
$$
is a lattice isomorphism, which means that it is one to one and
preserves the supremum and the infimum. Moreover,
$s \tr{i} t$ in $\bigsqcup_{n\geq 0} SPM(n)$ if
and only if $\chi(s) \tr{i+1} \chi(t)$ in \spmi.
\end{theo}
\proof
The application $\chi$ is obviously injective. Let us show that $\chi$ is surjective. Let
$s=(\infty, s_1,\ldots,s_l)$ be an element of \spmi.
Define $n=\sum_{i\geq 1} s_i$ and $s'=(s_1,\ldots,s_l)$. Since $s$ is in
\spmi, $s$ verifies the conditions of Theorem \ref{CaractElts}, and
so does $s'$. Therefore $s'$ is an element of \spmn,
and since $s=\chi(s')$, the application $\chi$ is surjective.

It is clear that  for all $s, t \in S$, one has $s \tr{i} t$ if and only if
$\chi(s) \tr{i+1} \chi(t)$. Obviously, $\chi$ is an order isomorphism.
Since \spmi\ is a lattice, we can conclude that
$\chi$ is a lattice isomorphism.
\cqfd

This result is illustrated in Figure \ref{fig_SPM_S} (left).
In the following, we simplify the notations by representing
the elements of \spmi\ without their first column.
Our aim is now to construct large parts of \spmi. A first
solution is to construct \spmn\ for large values of $n$.
However, this does not lead to filters (a \emph{filter}
of a lattice is a subset of this lattice closed for the
supremum) of \spmi. We will now define special filters of
\spmi\ and explain how we can construct them efficiently.
For a given $n$, let us denote by $SPM(\le n)$ the
set $$\bigsqcup_{0\le i\le n}SPM(i).$$ For example,
$SPM(\le 7)$ is shown in Figure \ref{fig_SPM_S}.
It is easy to see that $SPM(\le n)$ is a filter of \spmi\ 
for all $n$.
The infinite lattice \spmi\ can be regarded as a
limit of this sequence of posets. The results
presented in this section give us an efficient
method to construct $SPM(\le n)$ for all $n$
(see Algorithm \ref{Algo_infi} and Theorem \ref{TheoComplex}).
Moreover, we show another property of $SPM(\le n)$.

\begin{prop}
The poset $SPM(\le n)$ is a sublattice of \spmi\ for all $n$.
\end{prop}
\proof
To show the claim, it suffices to consider
$s \in SPM(k)$ and $ t\in SPM(l)$, with \mbox{$k \leq l \leq n$}, and show
that $\inf(s,t)$ and $\sup(s,t)$ (which are in \spmi\ since \spmi\ is a
lattice) are also in $SPM(\leq n)$. Let $u = \sup(s,t)$ in \spmi.
Since $u \geq s$ in \spmi, there exists
an integer $m \leq k$ such that $u \in SPM(m)$, therefore $u$ is in $SPM(\leq n)$.
Let now $u' = \inf(s,t)$ in \spmi. Let $s'$ be the partition obtained by addition of $l-k$ grains on the first column of $s$. Then, $s' \in SPM(l)$
and $\inf(s',t)$ in \spmi\ is also in $SPM(l)$.
So, $u =\inf(s,t)$ is also greater than or equal
to $\inf(s',t)$ in \spmi, hence $u \in SPM(\le l)$.
\cqfd

\begin{algorithm}
\SetVline
\In{an integer $n$}
\Out{$SPM(\leq n)$}
\Begin{
- Init $Result$ with $SPM(0)$\;
- \For{$1 \le i \le n$}{
        - Extract $SPM(i)$ from $Result$:\\
         Start from $(0)$\;
         Depthfirst search $Result$ to get the connex part containing $SPM(i)$\;
        - Link $SPM(i)$ to $Result$:\\
        \ForEach{$s \in SPM(i-1)$}{
       Link $s \in SPM(i-1)$ to $\inc{s}{1}\ in\ SPM(i)$ with an edge $\tr{1}$\;}}
- \Return{$Result$}
}
\caption{ {\sc Construction of $SPM(\le n)$} \label{Algo_infi}}
\end{algorithm}

\fig{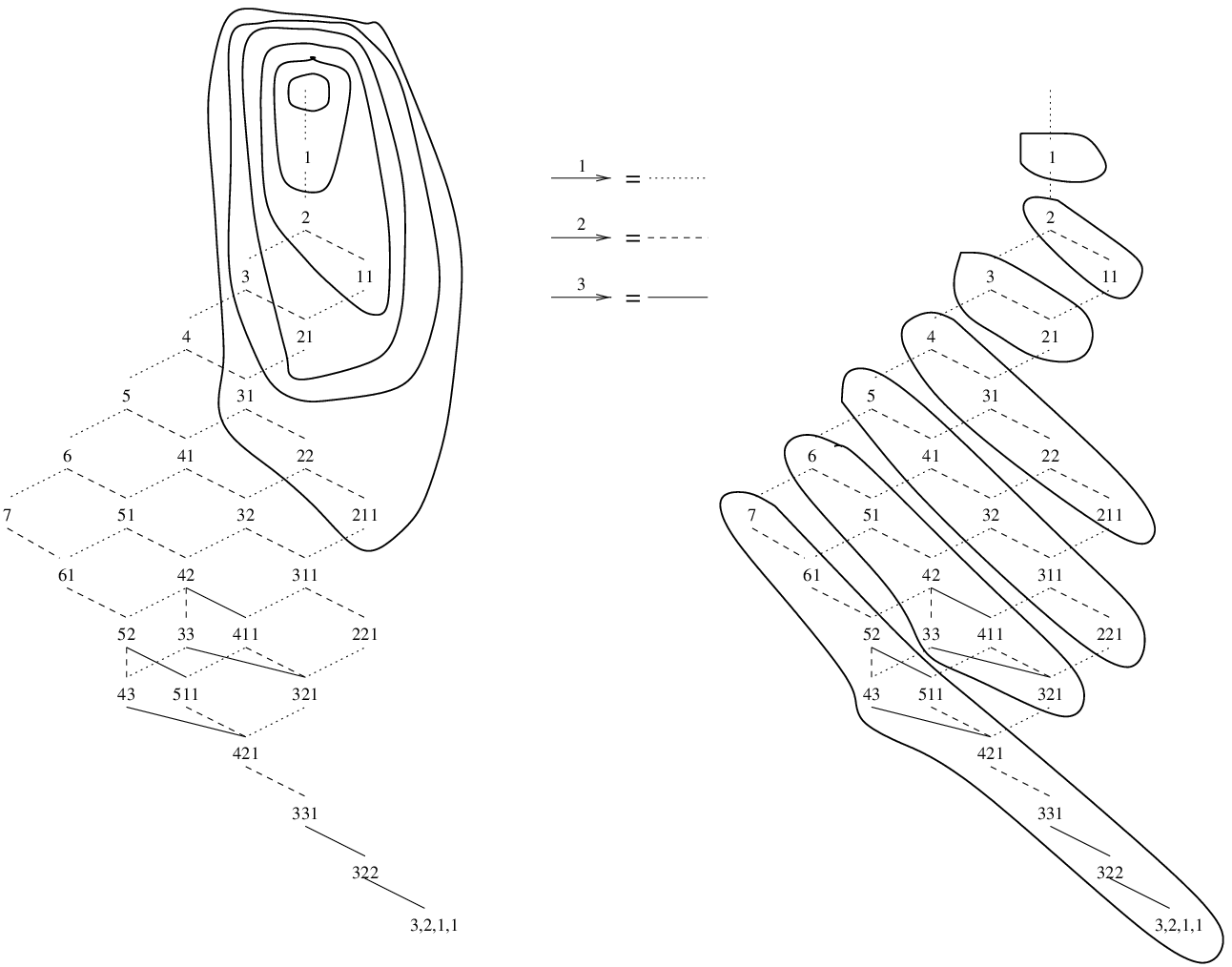}{The two ways to identify $SPM(n)$ in \spmi\ for all $n$.}

\begin{theo}
Algorithm \ref{Algo_infi} computes $SPM(\le n)$ in time linear
in the number of elements and edges in $SPM(\le n)$.
\label{TheoComplex}
\end{theo}
\proof
Let $|S|$ denote the number of elements and edges in $S$ for any lattice $S$.
For all $i$ between $1$ and $n$, the lattice $SPM(i)$ is a connected component
of $Result$. It is well known
that the computation of a connected component has a cost proportional to the
cardinality of this same component, so its extraction can be processed in
linear time.
The addition of the edges that link $SPM(i)$ to $SPM(i-1)$
can be done in
$O(|SPM(i-1)|)$. Therefore, each iteration of the \emph{for} loop is executed in
$O(|SPM(i)|)$, hence the execution of the whole loop is linear with respect to the total number of elements and
edges of $SPM(\le n)$. This is the asymptotic cost of the whole algorithm.
\cqfd

\subsection{The infinite tree \spt}

As shown in our construction of \spmnp\ from \spmn, each element
$s$ of \spmnp\ is obtained from an element $s' \in \spmn$ by
addition of one grain: $s = \inc{\mbox{$s'$}}{i}$ with $i$ an
integer between $1$ and $e(s')+1$. Thus, we can define an
infinite tree \spt\ (for Sand Pile Tree) whose nodes are
the elements of $\bigsqcup_{n\ge 0}{\spmn}$ and in which
the fatherhood relation is defined by:
$$
t \mbox{ is the $i$-th son of $s$ if and only if } t = \inc{s}{i}\ 
for\ some\ i,\ 1 \le i \le e(s)+1.
$$
The edge $s \longrightarrow \inc{s}{i}$ is labelled with $i$.
The root of this tree is $(0)$.
The eight first levels of \spt\ are shown in Figure~\ref{fig_arbre}
(we call the set of elements of depth $n$ the ``level $n$''
of the tree).
Each node $s$ of \spt\ has $e(s)+1$ sons linked to $s$ with edges
labelled \tr{1}, \tr{2}, $\dots$,
\raisebox{0.15cm}{$\underrightarrow{\mbox{\scriptsize $e(s)+1$ }}$}.

Notice that, although the notation is the same as the one used
for the \spm\ transitions in the lattice, an edge $s \tr{i} t$
in the tree means that $t$ is obtained from $s$ by addition of
one grain on its $i$-th column ($t = \inc{s}{i}$), and not that
$t$ is obtained from $s$ by having the top grain of the $i$-th
column fall onto the $(i+1)$-th. Therefore, if $s\in\spmn$
then $t\in\spmnp$.
So, the structure of the lattices is not directly
visible in \spt.
One goal in the following part of the section will be to explore
the possibility of the construction of the lattices from the tree.
The first results will be to state, as we could have guess from
Theorem \ref{th_chi}, that there are two ways to find \spmn\ in \spt.

\fig{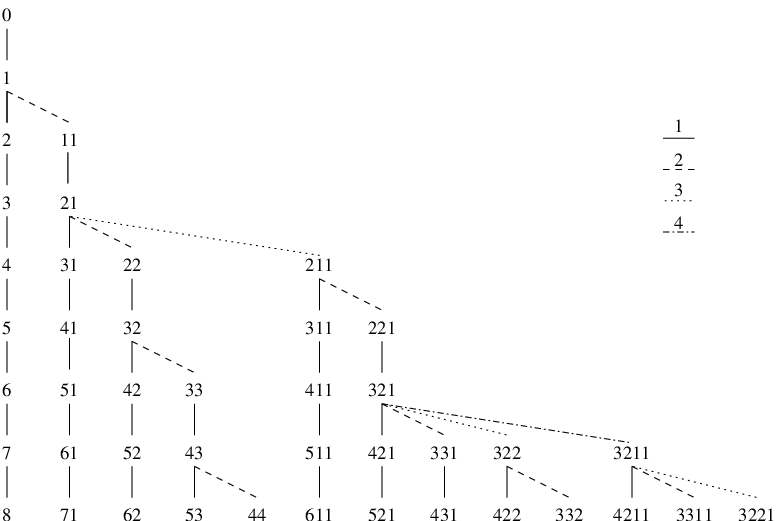}{The first levels of \spt.}

\begin{prop}
The level $n$ of \spt\ contains exactly the elements of \spmn.
\end{prop}
\proof
straightforward from the construction of \spmnp\ from \spmn\ given above.
\cqfd

\begin{prop}
For all integer $n$, the set $\overline{\spmn}$ is a subtree of \spt\ that contains its root.
\end{prop}
\proof
The proposition is obviously true for
\mbox{$n=1$}. Let us suppose it is true for $n$, and let us verify it is true
for $n+1$. By construction,
the elements of $\overline{SPM(n+1)} \setminus \overline{SPM(n)}$ are
sons of elements of $\overline{SPM(n)}$. The result follows.
\cqfd

We will now show that \spt\ can be described recursively, which allows us
to give a new recursive formula for $|\spmn|$.
Let us first consider one element $s$ of \spt\ and let
$k=e(s)$. By definition of \spt, $s$ has exactly $k+1$ sons.
Notice that the $k$ first sons of $s$ all verify the same
following property: for all integer $i$ between $1$ and $k$,
the son \inc{s}{i} begins with a stair of length $i-2$, has
a plateau at $i-1$, and a cliff at $i$. From this remark, we
introduce certain types of subtrees of \spt.
\begin{definition}
We call $N_k$ subtree, with $k\ge 1$, any subtree $T$ of \spt\ which
is rooted at an element $s$ that begins with a stair of length
$k-2$, has a plateau at $k-1$ and a cliff at $k$, and contains
all the descendants of $s$.
\end{definition}
\begin{definition}
We call $X_k$ subtree, with $k \ge 1$, any subtree $T$ of \spt\ which
is rooted at a node that has at least $k$ sons and such that the $i$-th
son is the root of a $N_i$ subtree for all $i$ between $1$ and $k$.
The structure of $X_k$ subtrees is shown in Figure~\ref{fig_def_Xk}.
Moreover, we define $X_0 = \emptyset$.
\end{definition}

\fig{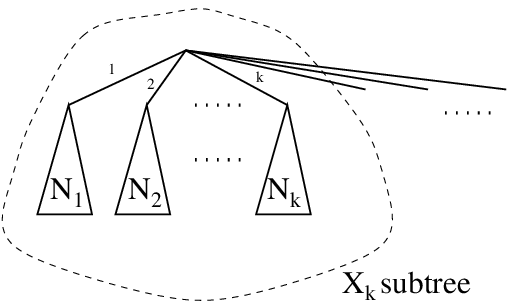}{Definition of the $X_k$ subtrees.}

Notice that \spt\ contains some $X_k$ subtrees. Indeed, if one
takes an element $s$ which begins with a stair of length $k$, then its $i$-th
son is the root of a $N_i$ subtree for all $i$, $1 \le i \le k$. So $s$ is the
root of a $X_k$ subtree. Notice also that $s$ might have other sons outside
the $X_k$ subtree.
With this remark, we characterise in the following proposition 
the structure of the $N_k$ subtrees, with $k \ge 1$.
See Figure \ref{fig_def_N1} and \ref{fig_Nk_struct}.

\begin{prop} We have the following statements~:
\begin{enumerate}
\item
A $N_1$ subtree is an infinite chain whose edges are all labelled $1$.
\item
A $N_k$ subtree, with $k\ge 2$, is composed by a chain of $k$ nodes whose
edges are labelled $k-1$, $k-2$, $\dots$, $1$ and whose $i$-th node is
the root of a $X_{k-1-i}$ subtree for all i between $1$ and $k-2$.
Moreover, the $k$-th node is root of a $X_k$ subtree.
\end{enumerate}
\end{prop}
\proof
\begin{enumerate}
\item
By definition, the root of a $N_1$ subtree is an element $s$ which begins
with a cliff, $e(s)+1 = 1$ (the conditions on the initial stair and the
plateau make no sense in this case). Therefore, the only son of $s$ is
$s \tr{1} \inc{s}{1}$, which also begins with a cliff. Therefore,
\inc{s}{1}\ is the root of a $N_1$ subtree and we can conclude by
induction.
\item
Let $k$ be greater than $1$.
Let us consider a partition $s$ such that:
\begin{itemize}
\item $e(s)=k-2$,
\item $s$ has a plateau at $k-1$,
\item and $s$ has a cliff at $k$.
\end{itemize}
The node $s$ has $k-1$ sons:
$s^{\downarrow_1}$, $s^{\downarrow_2}$, $\dots$, $s^{\downarrow_{k-1}}$.
Since $e(s)=k-2$, and from the remarks above,
the node $s$ is the root of a $X_{k-2}$ subtree that contains all
the elements reachable from its $k-2$ first sons.
Let $t$ be the $(k-1)$-th son of $s$ ($t$ is outside the $X_{k-2}$ subtree rooted
at $s$). The subtree rooted at $s$ is then the
union of a $X_{k-2}$ subtree and a subtree with root $t$
(see Figure \ref{fig_Nk_1}).
\fig{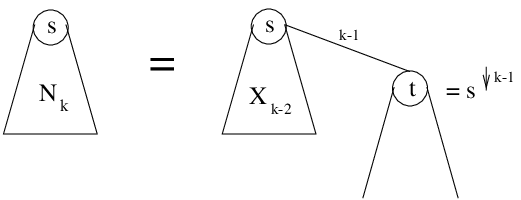}{First step of the structure of $N_k$.}
\noindent
Look now at the subtree with root $t$. We have that 
$e(t)=k-3$, hence $t$ is the root of a $X_{k-3}$ subtree. Let
$u$ be the $(k-2)$-th son of $t$. We then obtain Figure
\ref{fig_Nk_2}.
\fig{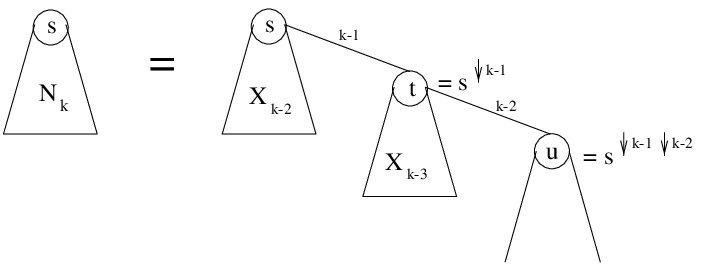}{Second step of the structure of $N_k$.}
When this process is iterated, we obtain
$x=s^{\downarrow_{k-1}\downarrow_{k-2}\ldots\downarrow_2}$. This element
$x$ begins with a plateau of length 1 followed by a stair of length
$k-2$ and a cliff. Therefore $x$ has only one son $x \tr{1} y$. This element
$y$ begins with a stair of length $k-1$ followed by a cliff. As noticed above,
$y$ is the root of a $X_k$ subtree and this subtree contains all the elements
reachable from $y$. Then we obtain the announced structure of $N_k$ subtrees,
see Figure \ref{fig_Nk_struct}.
\cqfd
\end{enumerate}

\fig{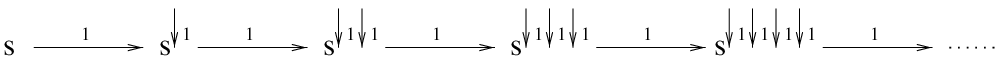}{Structure of the $N_1$ subtrees.}

\fig{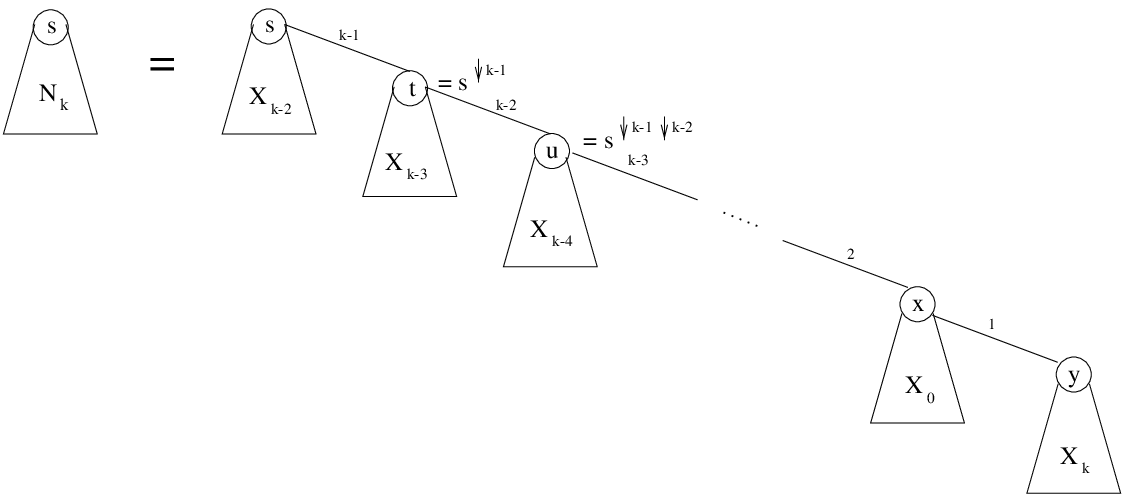}{Structure of $N_k$ subtrees.}

Using now the fact that a $X_k$ subtree is defined in terms of $N_k$ subtrees,
we can describe the structure of a $N_k$ subtree only in terms of other $N_i$
subtrees with $i \le k$ as shown Figure \ref{fig_Nk}.

\fig{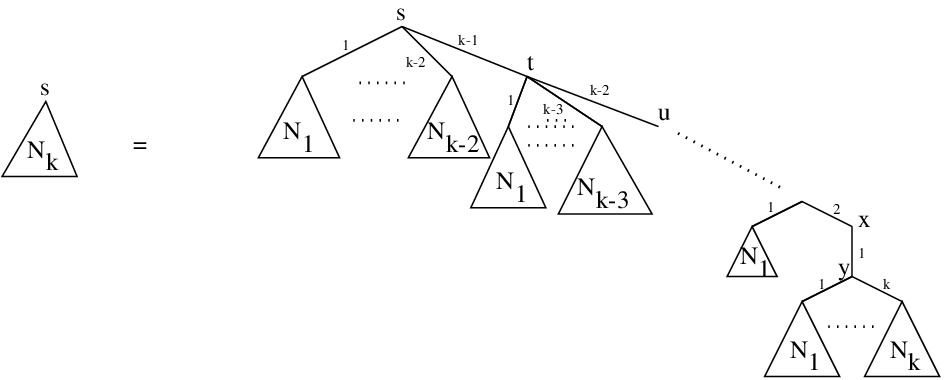}{Self referencing structure of $N_k$.}

Notice that we can deduce directly by induction
from the structure of the $N_k$ subtrees shown in
Figure \ref{fig_Nk} that all the
edges in a $N_k$ subtree are labelled with integer
smaller than or equal to $k$.
The recursive structures we have defined, and the
propositions given above allow a compact
representation of the tree \spt\ as a chain:
\begin{theo}
The tree \spt\ can be represented by the infinite chain shown in Figure \ref{fig_arbre_chaine}. The nodes of this chain are the fixed points of \spmn\ for $n\ge 0$.

The chain is defined as follows: let $k$ a positive integer and let $P_k = (k,k-1,k-2,\dots,2,1)$ and $P_{k+1} = (k+1,k,k-1,k-2,\dots,2,1)$; the subchain between $P_k$ and $P_{k+1}$ contains $k+1$ nodes:
$$
P_k\ \tr{k+1}\ \inc{P_k}{k+1}\ \tr{k}\ P_k^{\downarrow_{k+1}\downarrow_k}\ \tr{k-1}\ \ldots\ \tr{2}\ P_k^{\downarrow_{k+1}\dots\downarrow_2}\ \tr{1}\ P_{k+1}
$$
where each node $P_k^{\downarrow_{k+1}\dots\downarrow_i}$ with $i$ between
$3$ and $k+1$ is the root of a $X_{i-2}$ subtree, and $P_k$ is the root of
a $X_k$ subtree.
\end{theo}
\proof
Let us consider the rightmost chain in \spt.
This chain is composed by the fixed points of \spmn\ for $n \geq 0$. Let $k$
be a positive integer. Let us consider the subchain of this chain that begins
with $P_k= (k,k-1,\ldots,1)$ and terminates with
$P_{k+1}=(k+1,k,\ldots,1)$:
$$P_k \tr{k+1} P_k^{\downarrow_{k+1}} \tr{k} P_k^{\downarrow_{k+1}
\downarrow_k} \tr{k-1} \ldots \tr{2} P_k^{\downarrow_{k+1}\ldots\downarrow_2}
\tr{1} P_{k+1}.$$

The node $s = P_k^{\downarrow_{k+1}\ldots\downarrow_i}$ with $i\geq 2$, begins with
stairs of length $i-2$ followed by a plateau at $i-1$, hence $s$ is the root
of a $X_{i-2}$ subtree and its last son is obtained by
$P_k^{\downarrow_{k+1}\ldots\downarrow_i} \tr{i-1}
P_k^{\downarrow_{k+1}\ldots\downarrow_{i-1}}$. This is the next node in the
chain. Therefore \spt\ can be described as indicated.
\cqfd

\fig{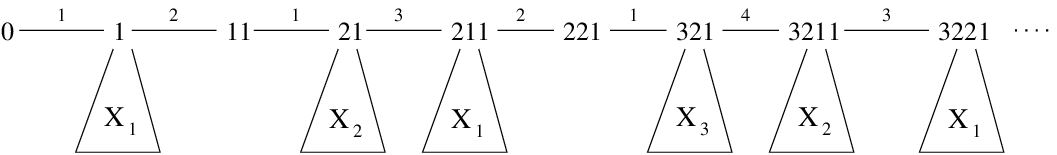}{The tree \spt\ represented as an infinite chain.}



As seen above, the level $n$ of \spt\ contains exactly \spmn.
Therefore, it suffices to count the number of paths of length
$n$ from the root of \spt\ to obtain $|\spmn|$. The recursive
structure of the tree, detailed above, gives us a way to
achieve this.
 
\begin{theo}
Let $c(l,k)$ denote the number of paths in a $X_k$ subtree originating from the root and of length $l$, then we have:
$$ c(l,k) = \left
\{\begin{array}{llll}
0  \quad \mbox{ if }  l\leq 0 \mbox{ and } k \leq 0 \\
1 \quad  \mbox{ if } l>0 \mbox{ and } k=1\\
k  \quad  \mbox{ if } l=1 \mbox{ and } k > 0\\
c(l-k,k) + \sum_{i=1}^{k-1} c(l-i+1,k-i) + \epsilon(l,k) &\mbox{ otherwise }
\end{array}
\right.$$
where $\epsilon(l,k) = 0$ if $k > l$ and $\epsilon(l,k) = 1$ otherwise.
\end{theo}
\proof
The proof follows from the recursive structure of $X_k$ detailed above.
There is no path of length $0$ or less, and the $X_k$ are empty for
$k\le 0$, hence the first case. A $X_1$ subtree is a simple chain, hence there is
exactly $1$ path of any length, hence the second case. The third case is
immediately deduced from the fact that the root of a $X_k$ subtree has exactly $k$ sons. Finally,
the recursive formula in the fourth case comes from the fact that the structure
of $X_k$ subtrees shown in Figure \ref{fig_def_Xk} allows us to consider a $X_k$
subtree as a node $s$ where $s$ is the root of a $X_{k-1}$ subtree and has one
more son which is the root of a $N_k$ subtree. Then, from the structure
of $N_k$ subtrees in terms of $X_k$ ones shown in Figure \ref{fig_Nk_struct}, we
deduce a description of $X_k$ subtrees in terms of $X_i$ subtrees with $0 \le i \le k$,
from which the formula is straightforward.
\cqfd

\begin{corol}
The cardinality of \spmn\ is given by:
$$
| SPM(n) | = 1 + \sum_{i=1}^{k} \sum_{j=1}^{i} c(n-\frac{i(i-1)}{2}-j+1, i-j+1)
$$
where $k$ is the integer such that
$\frac{k(k+1)}{2} \le n < \frac{(k+1)(k+2)}{2}$.
\end{corol}
\proof
This formula is deduced from the chain structure of the tree, shown in
Figure \ref{fig_arbre_chaine}. The quantity $1$ corresponds to the path of length $n$
that follows the chain without entering in a $X_i$ subtree. The double sum
corresponds to the repartition of the $X_i$ subtrees along the chain.
\cqfd



We will now show how information on \spmi\ can be deduced from \spt.
The lattice structure \spmi\ and the infinite tree \spt\ are defined
over the same underlying set: $\bigsqcup_{n\ge 0}\spmn$. Therefore
we can easily give a bijection from one to the other. We now show
how the ordered structure of \spmi\ can be deduced from \spt.

\begin{prop}
Every element of \spt\ has an outgoing edge with label $1$ in \spmi.
Moreover, for $i \ge 1$, a partition $s$ of \spt\ has an outgoing edge
\tr{i+1} in \spmi\ if and only if $s$ belongs to a $N_i$ subtree of
\spt. It should be noticed that $s$ may belong to several $N_i$
subtrees for dictinct values of $i$. In this case, $s$ is the
origin of an edge \tr{i+1} in \spmi\ for each such $i$.
In other words, $a$ is in a $N_i$ subtree of \spt\ if and only if
there is an outgoing edge from $a$ labelled with $i+1$ in \spmi.
\end{prop}
\proof
By definition of \spmi, every partition in \spmi\ obviously has a successor
by \tr{1}. Suppose now $i \ge 1$ and suppose that $a$ is
in a $N_i$ subtree of \spt, and denote by $s$ the root of this
$N_i$ subtree. By definition of the $N_i$ subtrees, we have that
$s_i - s_{i+1} \ge 2$, hence $a_i - a_{i+1} \ge 2$.
Therefore, when adding the first infinite column to obtain the corresponding
element of \spmi, we obtain that
there is an ouotgoing edge from $a$ labelled $i+1$ in \spmi.

Let us now consider a partition $a$ having an outgoing edge labelled
with $i+1$ in \spmi.
From the structure of the tree shown in Figure \ref{fig_arbre_chaine}, we know
that $a$ is in a $X_k$ subtree of \spt. Indeed, the elements of the chain are
fixed points of \spm, hence they have no outgoing edge in \spmi\ except the
ones labelled $1$, as said above.

Let $s$ be the root of such a $X_k$ subtree of \spt, that is, $s$ is the root
of a $X_k$ subtree that contains $a$. Consider a path from $s$ to $a$ in this
tree:
$$
s = s_0 \tr{i_1} s_1 \tr{i_2} s_2 \tr{i_3} \dots \tr{i_{l-1}} s_{l-1} \tr{i_l}
s_l = a.
$$
We have that $s_i - s_{i+1} \le 1$ by definition and $a_i - a_{i+1} \ge 2$
since there is a transition from $a$ labelled $i+1$ in \spmi. Therefore, there
exists an integer $j$ such that $i_j = i$. In fact, we have an even stronger
condition on the path: it must verify $|i+1| < |i|$ where $|x|$ denotes the
number of edges labelled with $x$ on the path.
But it is easy to see from the structure of the $N_i$ subtrees shown in Figure
\ref{fig_Nk} that the only case where this happends is when
$s_j \tr{i} s_{j+1}$ such that $s_{j+1}$ is the root of a $N_i$ subtree,
hence $a$ is in a $N_i$ subtree, as announced.
\cqfd

It follows from this proposition that we can find all the (immediate)
successors of a partition $s$ in \spmi: it is sufficient to go from
the root of \spt\ to $s \in \spt$ and so determinate the integers $i$
such that $s$ is in a $N_i$ subtree.

\section{Conclusion and Perspectives}

Through the study of the construction of \spmnp\ from \spmn,
we obtained much information about this set. First, it is
strongly self-similar and can be constructed using this property.
Moreover, this construction procedure gives a formula for
the cardinal of \spmn, where no formula was known before.
In a second part, we gave a natural way to extend \spmn\ to
infinity, and again self-similarity of this infinite lattice
appeared. Finally, we gave a tree structure to the sets \spmn\ and
\spmi, which allows efficient enumeration of \spmn, as well
as another formula for the cardinal of \spmn.

The duplication process that appears during the construction of
the lattices \spmn\ may be much more general, and
could be extended to other kinds of lattices, maybe leading to the
definition of a special class of lattices, which contains the lattices
\spmn. Moreover, the ideas developped in this paper
could be applied to others dynamical systems,
such as the Brylawski dynamical system \cite{Brylawski},
 Chip Firing Games, or tilings with flips, with some benefit.

\bibliographystyle{alpha}
\bibliography{SPM_infi}

\end{document}